\documentclass{article}

\usepackage{arxiv}

\usepackage{amsthm}
\usepackage{amsmath}
\usepackage[utf8]{inputenc} 
\usepackage[T1]{fontenc}    
\usepackage{hyperref}       
\usepackage{url}            
\usepackage{booktabs}       
\usepackage{amsfonts}       
\usepackage{nicefrac}       
\usepackage{microtype}      
\usepackage{lipsum}
\usepackage{graphicx}
\graphicspath{ {./images/} }
\usepackage[authoryear]{natbib}
\usepackage{pdfpages}
\theoremstyle{plain}

\newtheorem{thm}{Theorem}

\newtheorem{korollar}[thm]{Corollary}

\theoremstyle{definition}
\newtheorem{bemerkung}[thm]{Remark}
\newtheorem{example}[thm]{Example}
\newtheorem{definition}[thm]{Definition}

\usepackage[]{graphicx}
\usepackage{amsfonts} 
\usepackage{multirow}
\usepackage[dvipsnames]{xcolor}
\usepackage{booktabs}
\usepackage{siunitx}
\usepackage[section]{placeins}
\usepackage{subfigure}
\newcommand{\Blank}{\,\cdot\,}
 \DeclareMathOperator{\range}{range}
 \DeclareMathOperator{\Tr}{tr}

\title{Optimal designs for identifying effective doses in drug combination studies}

\author{
 Leonie Schürmeyer \\
  Department of Statistics\\
  TU Dortmund University\\
  Dortmund, 44227, Germany \\
  \texttt{schuermeyer@statistik.tu-dortmund.de} \\
   \And
 Ludger Sandig \\
  Department of Statistics\\
  TU Dortmund University\\
  Dortmund, 44227, Germany \\
  \And
  Jorrit Kühne \\
  Department of Statistics\\
  TU Dortmund University\\
  Dortmund, 44227, Germany \\
  \And
 Leonie Theresa Hezler \\
  Global Experimental Medicine, Data and Analytics\\
  Boehringer Ingelheim Pharma GmbH \& Co. KG\\
  Biberach an der Riß, 88397, Germany\\
  \And
  Bernd-Wolfgang Igl \\
  Global Experimental Medicine, Data and Analytics\\
  Boehringer Ingelheim Pharma GmbH \& Co. KG\\
  Biberach an der Riß, 88397, Germany \\
  \And
  Kirsten Schorning \\
  Department of Statistics\\
  TU Dortmund University\\
  Dortmund, 44227, Germany \\
  \texttt{schorning@statistik.tu-dortmund.de}\\
}

\begin{document}
\maketitle
\begin{abstract}
We consider the optimal design problem for identifying effective dose combinations within drug combination studies where the effect of the combination of two drugs is investigated. Drug combination studies are becoming increasingly important as they investigate potential interaction effects rather than the individual impacts of the drugs.
In this situation, identifying effective dose combinations that yield a prespecified effect is of special interest.
If nonlinear surface models are used to describe the dose combination-response relationship, these effective dose combinations result in specific contour lines of the fitted response model.

We propose a novel design criterion that targets the precise estimation of these effective dose combinations. In particular, an optimal design minimizes the width of the confidence band of the contour lines of interest. Optimal design theory is developed for this problem, including equivalence theorems and efficiency bounds. The performance of the optimal design is illustrated in different examples modeling dose combination data by various nonlinear surface models. It is demonstrated that the proposed optimal design for identifying effective dose combinations yields a more precise estimation of the effective dose combinations than ray or factorial designs, which are commonly used in practice. 
This particularly holds true for a case study motivated by data from an oncological dose combination study.

\end{abstract}


\section{Introduction}

In various disciplines of pharmaceutical development, analyzing potential drug-drug interactions is fundamental, and drug combination studies are becoming increasingly relevant (see e.g., \cite{chou2008preclinical} and \cite{mokhtari2017combination} among many others). In general, the detection of positive drug-drug interactions might result in an increased efficacy, reduced adverse events, and cost reduction. For this purpose, one focus of dose combination studies is on identifying dose combinations that achieve a prespecified effect.

\noindent

However, little research is available on modeling dose combination-response data and designing experiments for dose combinations in practice \citep{hollandletz2018}.

Often, the individual dose-response relationships of the monotherapies are known, but their combination is of particular interest, especially when the mode of action is unknown. In this situation, there are different approaches to specify the type of interaction, i.e. whether there is synergism, additivity, or antagonism of the investigated drugs, which correspond to a positive interaction, no interaction, or a negative interaction, respectively (see e.g., \cite{loewe1927mischarznei}, \cite{bliss1939toxicity} and \cite{lee2007interaction}). Here, common strategies include, for example, curve-shift analysis, isobolograms, combination indices, and universal surface response analysis (see, e.g., \cite{zhao2010comparison} and \cite{foucquier2015analysis}). While modeling the drug-drug interaction based on the combination index or isobolograms results in a one-dimensional linkage not capturing the response effect itself at each possible dose combination, dose response surface modeling enables an estimation of the drug-drug interaction across the entire design space based on the response effect. Therefore, we consider the situation in which the drug-drug interaction is modeled by dose response surface models. In particular, we consider two different specific approaches for response surface modeling. One is the Greco model, which is a commonly used modification of the sigmoid Emax model, often employed in one-dimensional settings (see, e.g., for the sigmoid Emax \cite{macdougall2006analysis} and \cite{liou2015response} for the Greco model). The second model, proposed by, \cite{zhou2024combination} uses the established one-dimensional dose-responses models in an additive manner, and includes an interaction effect resulting in specific nonlinear surface models to model the dose combination response data.
In general, the two-dimensional modeling by response surface models introduces new challenges for the design of the corresponding experiments. Similar to traditional one-dimensional dose-response experiments, researchers are interested not only in modeling the effects across the entire design space but also in evaluating specific effective doses, which are defined as the dose combinations that achieve a predetermined target effect size.
In the one-dimensional setting, the dose that results in a specified effect level is unique, particularly if the response is strictly increasing with dose. In the two-dimensional setting, multiple dose combinations yield a specific effect level, forming contour lines on the surface model, even if the response is strictly increasing in both doses. Therefore, a design in drug combination studies that facilitates precise estimation of these effective dose combinations leading to a prespecified effect level of interest is needed.
\\
This paper aims to construct efficient designs for achieving a target effect size using nonlinear surface models. Although the theoretical and practical considerations of optimal designs for different kinds of classical one-dimensional dose-response models are well established (see, e.g., \cite{bretz2010practical}, \cite{dette2016optimal} and \cite{bornkamp2009mcpmod}), there is less research on the design of drug combination experiments. Ray designs, whose dose levels correspond to fixed proportions of the mixture of substances, are frequently used in drug combination studies (see e.g., \cite{straetemans2005design} and \cite{ronneberg2021bayesynergy}). Therefore, \cite{almohaimeed2014experimental} established locally D-optimal designs for specific ray designs.
\cite{hollandletz2018} consider optimal experimental designs for estimating the drug combination index based on Loewe additivity. Besides, \cite{papathanasiou2019optimizing} consider nonlinear surface models to describe the dose combination response relationship across the whole design space but mainly address the design problem by using the classical D-optimality criterion instead of explicitly targeting the precise estimation of effective dose combinations. To address this gap, we propose a new and flexible design criterion that aims to accurately estimate effective dose combinations in drug combination studies. In particular, the design criterion can be used in combination with a broad class of (non) linear response surface models such as the Greco-model or the additive sigmoid Emax model.
\\
 The paper is structured as follows. First, we introduce a definition of effective dose combinations under the assumption of a (non) linear response surface model in Section \ref{sec:Methodology}. Here,  we also present two specific approaches of surface modeling as commonly used in drug combination studies. In Section \ref{sec:OptDes}, we will introduce the optimal design theory in the context of drug-combination studies aiming to estimate the effective dose combinations accurately. More precisely, the new developed optimal design criterion minimizes the $L_q$-norm of the variance of the prediction for the effect at the effective dose-combinations. Note that a similar approach is proposed by \cite{miller2007optimal} for the one dimensional set-up, but can not be directly extended to the present situation of dose combination studies.
For the present context, we will provide equivalence theorems and a lower bound for the efficiencies that can even be used if the optimal design is unknown.\\
In practice, the derived optimal designs should result in more precise predictions of the effective dose-combinations. Therefore, Section \ref{Case study} and Section \ref{sec:Num} are devoted to illustrating the advantages of the derived optimal designs in specific scenarios of dose combination studies. Based on a real case study, we will compare the performance of the derived optimal designs against established designs used in practice such as factorial and ray designs and classical established approaches from optimal design theory like the locally D-optimal design. The considered design approaches will be compared theoretically, based on efficiencies, and practically through a simulation study. In addition to the case study, another scenario is investigated, both reflecting practical scenarios of dose combination studies with different underlying types of surface models.\\
Finally, we examine the robustness of the developed design approach. In drug combination studies, monotherapies are often known in advance, while the interaction effect may not be clearly defined before conducting the drug combination experiment. Therefore, we consider robust designs for various values of the interaction effect. Additionally, we investigate the capability of the resulting designs to provide precise estimates of the effective doses and assess their performance under these varying conditions.

\section{Modeling dose combination data}\label{sec:Methodology}
We consider the effect of the combination of two different substances, denoted by $C$ and $D$, by modeling it with the nonlinear response surface model
\begin{equation}\label{eq:nonlin_surface}
  Y_{ij} = \eta((c_i,d_i),\theta) + \varepsilon_{ij}
  \quad\text{for  } i=1,\ldots,n; j=1,\ldots,r_i,
\end{equation}
where $\varepsilon_{ij}$ are independent random variables such that $\varepsilon_{ij} \sim \mathcal{N}(0,\sigma^2)$, $\sigma^2 >0$.
This means that observations are taken at $n$ different dose combinations $(c_1,d_1), \ldots, (c_n, d_n)$, which vary in the design space (i.e., the dose combination space) $\mathcal{Z} = \mathcal{X}_C \times \mathcal{X}_D = \left[ 0,c_{\max}\right]\times\left[ 0, d_{\max}\right] \subset \mathbb{R}^2$ and $r_i$ observations are taken at each $(c_i, d_i)$, $i=1, \ldots, n$. Let $N= \sum_{i=1}^{n}r_i$ denote the total sample size. The regression model~$\eta$ and the $m$-dimensional parameter $\theta$ are used to describe the relationship between the response and the dose combination. We assume that the function $((c,d), \theta) \mapsto \eta((c,d), \theta)$ is continuously differentiable both in~$(c,d)$ and in $\theta$.

\medskip

There are various approaches for modeling the combination of two substances differing by the choice of the regression function $\eta$. One natural choice for the regression function $\eta$ includes the regression functions used to describe the dose-specific effect of the individual substances. More precisely, let $\eta_C(\Blank, \theta_C)$, $\theta_C\in \Theta_C \subset \mathbb{R}^{m_C}$, and $\eta_D(\Blank, \theta_D)$, $\theta_D\in \Theta_D \subset \mathbb{R}^{m_D}$ be the parts of the individual regression functions describing the dose-specific effect of the substances $C$ and $D$, respectively. Then we define the corresponding regression function for the dose combination in an additive manner including an interaction effect with
\begin{equation}
    \label{interactmod}
    \eta((c, d), \theta) = \theta_0 + \eta_C(c, \theta_C) + \eta_D(d, \theta_D) + \gamma \eta_C(c,\theta_C)\eta_D(d, \theta_D)\, ,
\end{equation}
where the parameter $\theta= (\theta_0, \theta_C, \theta_D, \gamma)$ consists of the individual parameters $\theta_C$ and $\theta_D$ of the regression functions $\eta_C$ and $\eta_D$, a joint placebo parameter $\theta_0$ and the parameter $\gamma \in \mathbb{R}$ that describes potential dose-independent interaction effects between the two substances. Note, that due to identifiability reasons the parameters $\theta_C$ and $\theta_D$ do not contain a substance specific parameter for the placebo-effect. The interaction effect is positive if $\gamma > 0$, whereas $\gamma < 0$ indicates a negative interaction. If $\gamma = 0$, there is no interaction between the two substances. A drug-drug interaction may enhance the efficacy of a treatment even at lower dose ranges, which could additionally lead to a reduction in the occurrence of adverse events, for example, in oncology studies. Summarizing, for the regression function $\eta$ in \eqref{interactmod} it holds for the dimension of the parameter $\theta$ that $m= 2+ m_C+m_D$. In clinical and toxicological applications popular choices for the function $\eta_C$ and $\eta_D$  are the linear, exponential, Emax, or sigmoid Emax model, as described by \cite{bornkamp2009mcpmod} and presented in Table \ref{tab:ModelChoices}. Note that the functions depicted in Table \ref{tab:ModelChoices} are continuously differentiable, hence also $\eta$ is differentiable.

\begin{table}
    \centering
    \caption{Possible choices of the one-dimensional regression functions for $\eta_C$ and $\eta_D$ in the dose combination model \eqref{interactmod}. Note that the placebo parameter of the depicted regression functions from \cite{bornkamp2009mcpmod} is removed, as it is already part of the dose combination model \eqref{interactmod}.}
     \label{tab:ModelChoices}
\begin{tabular}{lll}
 \toprule
 Model & Formula & Parameters \\
 \midrule
 Linear & $f(d,\theta_f)=\delta d$ & $\theta_f=\delta$\\[0.2cm]
 Exponential & $f(d,\theta_f)=\mathrm{E}_1(\exp(\frac{d}{\delta})-1)$ & $\theta_f= (\mathrm{E}_1,\delta)$\\[0.2cm]
 Emax & $f(d,\theta_f)=\mathrm{E}_{\max} \frac{d}{\mathrm{ED}_{50}+d}$ & $\theta_f=(\mathrm{E}_{\max}, \mathrm{ED}_{50})$ \\[0.2cm]
 sigmoid Emax & $f(d,\theta_f)=\mathrm{E}_{\max} \frac{d^h}{\mathrm{ED}_{50}^h+d^h}$ & $\theta_f=(\mathrm{E}_{\max}, \mathrm{ED}_{50} , h)$\\[0.2cm]
 \bottomrule
\end{tabular}
\end{table}

Another natural choice for the regression function $\eta$ is included in the response surface model called Greco model. According to \cite{liou2015response} the regression function $\eta$ of the Greco model is defined by
\begin{align}
\label{GrecoMod}
    \eta((c, d), \theta) = \frac{\mathrm{E}_{\max} \cdot \bigl[ \frac{c}{\mathrm{ED}_{50\, C}} + \frac{d}{\mathrm{ED}_{50\, D}} + \alpha \cdot (\frac{c}{\mathrm{ED}_{50\, C}}\cdot \frac{d}{\mathrm{ED}_{50\, D}}) \bigr]^\gamma }{\bigl[ \frac{c}{\mathrm{ED}_{50\, C}} + \frac{d}{\mathrm{ED}_{50\, D}} + \alpha \cdot (\frac{c}{\mathrm{ED}_{50\, C}}\cdot \frac{d}{\mathrm{ED}_{50\, D}}) \bigr]^\gamma + 1},
\end{align}
where the parameter $\theta=(\mathrm{E}_{\max}, \mathrm{ED}_{50\, C}, \mathrm{ED}_{50\, D}, \alpha, \gamma)$ consists of the individual model parameters. In detail, $\mathrm{E}_{\max}$ is the maximal achievable effect, which has to be the same for both substances. $\mathrm{ED}_{50\, C}$ and $\mathrm{ED}_{50\, D}$ describe the effective doses, where substances $C$ and $D$ reach 50\% of the maximal effect, respectively, when there are given alone. Besides, $\alpha$ refers to the interaction effect. A synergistic effect is observed for $\alpha>0$, while $\alpha <0$ indicates an infra-additive effect. An additive interaction occurs for $\alpha =0$. The steepness of the surface model is indicated by $\gamma$. 
The model can also be extended with an intercept by adding the placebo effect $\mathrm{E}_0$ to \eqref{GrecoMod}.

  The structure of the Greco model is based on the one-dimensional sigmoid Emax model (see Table \ref{tab:ModelChoices}) and can be seen as a modification of this.
  However, note the different levels at which the interaction is incorporated in the two models:
  in~\eqref{interactmod}, the interaction combines the responses for the different substances,
  while~\eqref{GrecoMod}, the interaction happens already at the level of the concentrations.

Using the model defined in \eqref{eq:nonlin_surface} with a specified regression function $\eta$, we assume for asymptotic arguments that $\lim_{N \rightarrow \infty} \tfrac{r_i}{N} = \omega_i \in (0,1)$ and collect this information in the matrix
\begin{align*}
  \xi = \begin{pmatrix}
    (c_1,d_1) & \cdots & (c_n,d_n)\\
    \omega_1 & \cdots & \omega_n
  \end{pmatrix}.
\end{align*}
Following \cite{kiefer}, we refer to $\xi$ as an approximate design on the design space $\mathcal{Z}$. This means that the dose combinations $(c_i, d_i)$ define the different experimental conditions where observations are to be taken, and the weights $\omega_i$ represent the relative proportion of observations at the corresponding dose combination $(c_i, d_i)$. If an approximate design is given and $N$ observations can be taken, a rounding procedure proposed by \cite{pukelsheimrieder1992} is applied to obtain integers $r_i$ from the not necessarily integer valued quantities $N\omega_i$.
Assume that observations are taken according to an approximate design $\xi$.
Under certain assumptions of regularity, the distribution of $\hat{\theta}$ is asymptotically normal \citep{jennrich1969asymptotic}. Moreover, the distribution of the predicted effect at a specific dose combination $(c_0, d_0)$ is also asymptotically normal, with
\begin{equation*}
  \sqrt{N}\bigl(\eta\bigl((c_0,d_0),\hat{\theta}\bigr) - \eta\bigl((c_0,d_0),\theta\bigr)\bigr)
  \overset{\mathcal{D}}{\rightarrow}
  \mathcal{N}\bigl(0,\varphi \bigl((c_0,d_0),\xi,\theta\bigr)\bigr).
\end{equation*}
Here, $\overset{\mathcal{D}}{\rightarrow}$ denotes convergence in distribution and the function $\varphi$ is defined by
\begin{align}
\label{asymptotic_pred}
  & \varphi \bigl((c_0,d_0),\xi,\theta\bigr) = \left(\frac{\partial}{\partial \theta} \eta\bigl((c_0,d_0),\theta\bigr)\right)^T M^{-}(\xi,\theta)\left(\frac{\partial}{\partial \theta} \eta\bigl((c_0,d_0),\theta\bigr)\right) ,\\
  \label{infomat}
  \text{where}\quad  & M(\xi,\theta)= \int_{Z} \frac{\partial}{\partial \theta} \eta((c,d),\theta) \left(\frac{\partial}{\partial \theta} \eta((c,d),\theta)\right)^T d\xi((c,d))
\end{align}
is the information matrix corresponding to the design $\xi$, and $\tfrac{\partial}{\partial \theta} \eta((c,d),\theta)$ is the gradient of $\eta$ with respect to the parameter $\theta\in \mathbb{R}^m$. The matrix $M^{-}(\xi, \theta)$ is the generalized inverse of the information matrix, whereby the design $\xi$ has to satisfy $\tfrac{\partial}{\partial \theta} \eta((c_0,d_0),\theta) \in \range(M(\xi, \theta))$. \\
Therefore, the asymptotic variance of the prediction $\eta((c_0, d_0), \hat\theta)$ at a prespecified dose combination is given by $\varphi((c_0, d_0), \xi, \theta)$. The asymptotic behavior in \eqref{asymptotic_pred} can be used to construct an asymptotic confidence interval for the prediction of the effect at the dose combination $(c_0, d_0)$. More precisely, the confidence interval for the level $(1-\alpha)$ is given by
\begin{equation}
\label{pred_interval}
    \eta\big((c_0, d_0), \hat\theta\big) \pm  z_{1-\alpha/2} \frac{\hat\sigma}{\sqrt{N}}\varphi^{1/2}\big((c_0, d_0), \hat\theta\big) \, ,
\end{equation}
where $\hat\sigma$ and $\hat\theta$ denote the maximum likelihood estimates of the parameters $\sigma$ and $\theta$, $z_{1-\alpha/2}$ denotes the $(1-\alpha/2)$-quantile of the standard normal distribution and the function $\varphi$ is given in \eqref{asymptotic_pred}.

\subsection{Effective doses in dose combination studies}
In classical dose-response experiments where the effect of one substance is investigated, special interest lies on effective doses~(ED) that result in a certain percentage of the maximal effect. In particular, following \cite{bretz2010practical}, the $\mbox{ED}_p$-dose at which $p\%$ ($0 < p < 100$) of the maximal effect is achieved is defined by
\begin{equation}
  \label{one_dim_EDP1}
  \mbox{ED}_p= \min \biggl\{ x\in(a,b]\, \biggm| \frac{h(x, \theta)}{h(b, \theta)} = \frac{p}{100}\biggr\},
\end{equation}
where $h(x, \theta) = f(x, \theta) - f(0, \theta)$ and $f(x, \theta)$ is the considered regression function, which depends on both the dose $x$ and the unknown parameter $\theta$.
Note that using the definition in \eqref{one_dim_EDP1} results in a unique $\mbox{ED}_p$.
If the regression function $f(x, \theta)$ is additionally strictly increasing, the effect level $p$ is attained at exactly one dose.

However, in the two-dimensional setting of dose combination studies, there is no unique $\mbox{ED}_p$-dose combination, even if the regression function $\eta((c, d), \theta)$ is strictly increasing in both doses $c$ and $d$. Instead, several different dose combinations $(c,d)\in \mathcal{Z}$ yield the same effect, resulting in contour lines on the surface model.
Extending the one-dimensional definition of the $\mbox{ED}_p$ in \eqref{one_dim_EDP1}, the set of multivariate effective doses for achieving $p\%$ ($0 < p < 100$) of the maximum effect in the considered design range $\mathcal{Z}=\left[0,c_{\max}\right]\times \left[0, d_{\max}\right]$ can be defined by
\begin{equation}\label{MED_set}
    \mathrm{MED}_p(\theta) = \biggl\{ (c,d) \in \mathcal{Z} \,\biggm|\,
    \frac{\eta((c,d),\theta) -\min_{(c_0,d_0)\in\mathcal{Z}}\eta((c_0,d_0),\theta)}{R_{\max} }
    = \frac{p}{100}  \biggr\},
\end{equation}
where
the $R_{\max}$ denotes the maximal effect of the response in the design space and is given by
\begin{equation*}
  R_{\max}=\max_{(c,d)\in\mathcal{Z}}\eta((c,d),\theta) - \min_{(c,d)\in\mathcal{Z}}\eta((c,d),\theta).
\end{equation*}
Note that the $\mathrm{MED}_p(\theta)$ defined in \eqref{MED_set} depends on the parameter $\theta$. For the sake of brevity, we denote the multivariate effective doses for $p$ \% of the maximal effect with $\mathrm{MED}_p$.

\begin{figure}
  \centering
  \subfigure[Surface model with contour lines $p_1=80$, $p_2=90$.]{\includegraphics[width=0.49\textwidth]{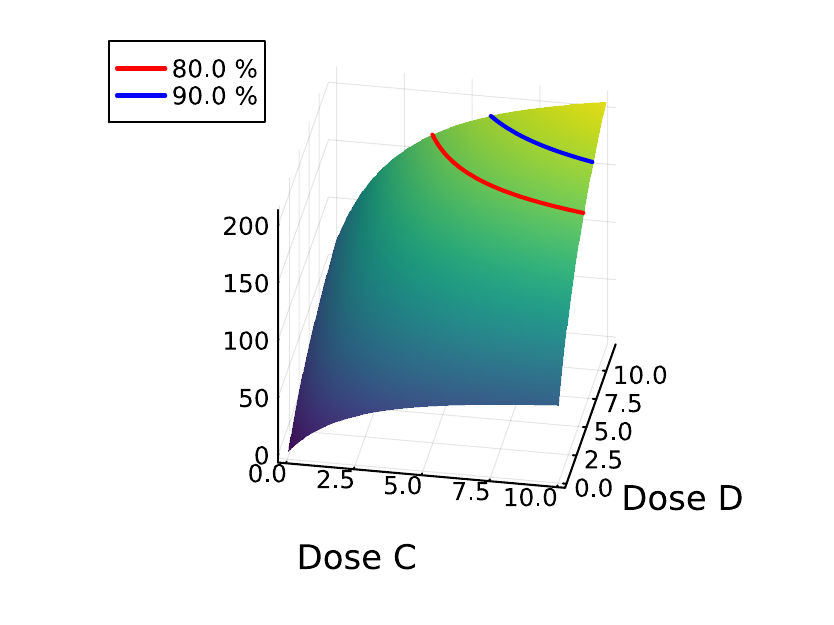}}
  \subfigure[Visualization of the $\mathrm{MED}_{80}$ and $\mathrm{MED}_{90}$.]{\includegraphics[width=0.49\textwidth]{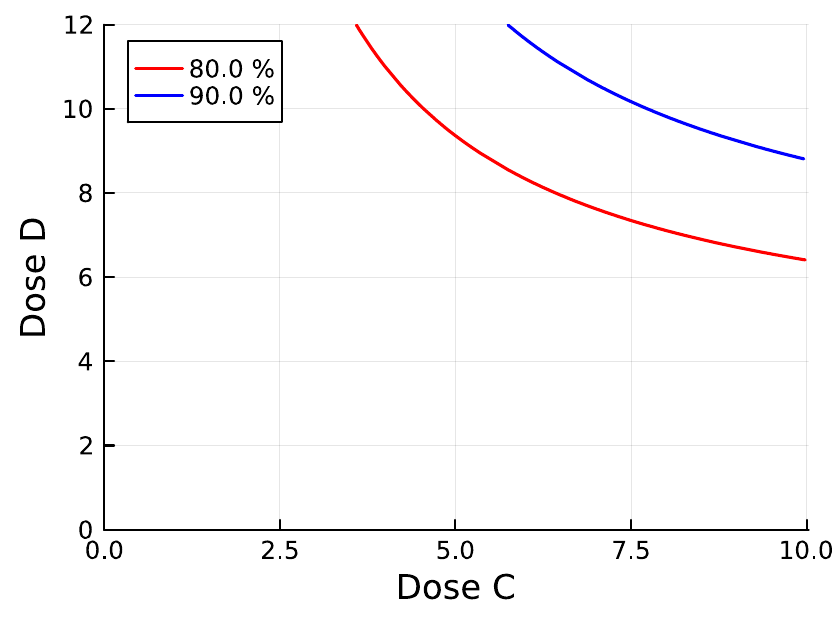}}
  \caption{Response surface and multivariate effective doses for the model from Example~\ref{example-model}.}
  \label{fig: ExampleMED}
\end{figure}

\begin{example}
  \label{example-model}
Consider the situation where dose combinations of the substances $C$ and $D$ that achieve $p_1=80\%$ and $p_2=90\%$ of the maximal effect are of interest in a dose combination study. Further, assume that the dose combination response relationship can be described by an additive interaction model \eqref{interactmod}, where the individual regression functions are given by Emax models with parameter $\theta_C= (80,3)$ and $\theta_D=(120, 10)$, respectively.  The interaction effect is given by $\gamma=0.02$. The corresponding combination response surface model is shown in Figure \ref{fig: ExampleMED} (a). The dose combinations that result in $80\%$ and $90\%$ of the maximal effect, thus the sets $\mathrm{MED}_{80}$ and $\mathrm{MED}_{90}$, are marked by the red and blue (contour) line, respectively (Figure \ref{fig: ExampleMED} (b)).
\end{example}

\section{Optimal design theory for identifying sets of effective doses in drug combination studies}\label{sec:OptDes}

Dose combination studies aim to identify the dose combinations $(c,d) \in \mathrm{MED}_p$ for prespecified values for $p\%$. Using \eqref{pred_interval} the pointwise confidence band of the prediction of effects at the dose combinations contained in the $\mathrm{MED}_p$ is given by
\begin{equation}
\label{pred_band}
    \biggl\{\eta((c, d), \hat\theta) \pm  z_{1-\alpha/2} \frac{\hat\sigma}{\sqrt{N}}\varphi^{1/2}((c, d), \xi, \hat\theta) \, \biggm|\, (c, d) \in \mathrm{MED}_p\biggr\},
\end{equation}
where the asymptotic variance function $\varphi((c, d), \xi, \hat\theta)$ is defined in \eqref{asymptotic_pred}. The smaller the width of this confidence band is, the more precise the prediction of the effect at dose combinations $(c,d)$ contained in $\mathrm{MED}_p$ becomes. A precise prediction of the effect at those dose combinations, in turn, leads to a precise identification of the corresponding $\mathrm{MED}_p$. Note that, by definition, it is not possible to construct a criterion which aims at a precise estimation of the MEDs directly, as these depend on the estimated curve itself. Due to this, the precision of the MEDs needs to be assessed via the precision of the estimated effects at the corresponding dose combinations. Consequently, from a design perspective, a good design $\xi$ should minimize the width of the confidence band in \eqref{pred_band} at each dose combination $(c, d) \in \mathrm{MED}_p$. This corresponds to a minimization of the asymptotic variance in \eqref{asymptotic_pred} concerning the design $\xi$. Unfortunately, a simultaneous minimization is only possible in rare and probably unrealistic settings. We therefore propose a design criterion to minimize a $L_q$-norm of the function $\varphi$, $q\in [1, \infty)$.  More precisely, let $p_1, \ldots, p_k\in (0,100)$ be $k\geq 1$ prespecified percentages and $\mathcal{C}(\theta)=\bigcup_{i=1}^k \mathrm{MED}_{p_i}\neq \emptyset$ the joint set of the corresponding MED-sets. Then, we can use the $L_q$-norm
\begin{equation}
  \label{MED-crit}
  \phi_{\mathrm{MED}_q}(\xi,\theta)
  = \left(\frac{1}{l_{\mathcal{C}(\theta)}}
    \int_{\mathcal{C}(\theta)} \varphi^q \left((c,d),\xi,\theta\right)d\mu(c,d)
  \right)^{\frac{1}{q}}
\end{equation}
of the function $\varphi$ defined in \eqref{asymptotic_pred}, where $\mu$ denotes an appropriate measure on the design space $\mathcal{Z}$ and $l_\mathcal{C(\theta)}= \mu(C(\theta)) > 0$ the corresponding value of the measure $\mu$ evaluated at the set $C(\theta)$. Typical choices for the measure $\mu$ are the uniform distribution on the set $C(\theta)$ or discrete measures on the design space $\mathcal{Z}$ (as later chosen in the simulation study in Section \ref{SectionSimulationStudy}).
Note that the function in~\eqref{MED-crit} is similar to the I-criterion that is discussed in~\cite[p.~58]{fedorov2013optimal}.

\begin{definition}
    Let $p_1, \ldots, p_k \in (0,100)$, $k\geq 1$, $q\in[1, \infty)$.
    Then a design $\xi^{\ast}$ is called locally $\mathrm{MED}_q$-optimal for the $k$ effective dose combinations $\mathrm{MED}_{p_1},\ldots,\mathrm{MED}_{p_k}$ and the parameter $\theta$, if it
    satisfies $\tfrac{\partial}{\partial \theta} \eta((c,d),\theta) \in \range(M(\xi^\ast, \theta))$ for all $(c,d) \in \bigcup_{i=1}^k \mathrm{MED}_{p_i}$ and if it minimizes the function $\phi_{\mathrm{MED}_q}(\xi,\theta)$ in \eqref{MED-crit} over the space of all approximate designs $\xi$ on $\mathcal{Z}$ with $\tfrac{\partial}{\partial \theta} \eta((c,d),\theta) \in \range(M(\xi, \theta)$ for all $(c,d) \in \bigcup_{i=1}^k \mathrm{MED}_{p_i}$.

\end{definition}

A central tool of optimal design theory is the equivalence theorem, which is frequently used to check the optimality of candidate designs determined numerically. Popular algorithmic approaches are versions of Fedorov's exchange algorithm (e.g., \cite{yang-2013-optim-desig}), but also stochastic particle-base heuristics that do not have convergence guarantees (e.g., \cite{masoudi2019metaheuristic} or \cite{chen-2022-partic-swarm}). Moreover, equivalence theorems can be used to reduce the infinite dimensional optimization problems arising in optimal design theory to finite dimensional ones. Due to the convexity of the $L_q$-norm, the criterion derived in $\eqref{MED-crit}$ is convex with respect to the design $\xi$. Thus, we
can derive the corresponding equivalence theorem (Theorem \ref{equivtheorem}) that can be used to check the $\mathrm{MED}_q$-optimality of a given design $\xi^*$. Note that Theorem \ref{equivtheorem} consists of two parts. The first part is derived for the case where the information matrix $M(\xi^*, \theta)$ of the design $\xi^*$ is non-singular. However, there might be cases where the information matrix $M(\xi^*, \theta)$ of the ${\mathrm{MED}_q}$-optimal design $\xi^*$ is singular. For instance, this situation can arise if only $k=1$ set of effective concentrations, denoted as $\mathrm{MED}_p$, is of interest. In Section \ref{Case study}, the statement of the theorem is used to check the optimality of the numerically determined designs. The proof of Theorem \ref{equivtheorem} can be found in the Supplementary Material.

\begin{thm}
  \label{equivtheorem}
  \begin{enumerate}
  \item Let $\xi^*$ be an approximate design on $\mathcal{Z}$ such that the corresponding information matrix $M(\xi^*, \theta)$ is non-singular. The design $\xi^{\ast}$ is locally $\mathrm{MED}_q$-optimal if and only if the inequality
    \begin{align}
      \label{EquTheoremMED1}
      &\int_{\mathcal{C}(\theta)} \bigl( \varphi((c,d),\xi^\ast,\theta)\bigr)^{q-1}
        \alpha^2\bigl((c_0,d_0), (c,d),\xi^\ast,\theta\bigr)\,d \mu(c,d) - \phi_{\mathrm{MED}_q}^q(\xi^\ast,\theta)\leq 0 \\
      &\text{with  } \alpha\bigl((c_0,d_0),(c,d),\xi,\theta\bigr)= \left(\frac{\partial}{\partial \theta} \eta((c,d),\theta)\right)^T M^{-1}(\xi,\theta)\left(\frac{\partial}{\partial \theta} \eta((c_0,d_0),\theta)\right),\nonumber
    \end{align}
    holds for all $(c_0,d_0)\in\mathcal{Z}$.
    Moreover, equality is achieved in \eqref{EquTheoremMED1} for all $(c,d)$ in the support of the design $\xi^*$.
  \item Let $\xi^*$ be an approximate design on $\mathcal{Z}$ such that the corresponding information matrix $M(\xi^*, \theta)$ is singular. The design $\xi^{\ast}$ is locally $\mathrm{MED}_q$-optimal if and only if there exists a generalized inverse $G\in \mathbb{R}^{m\times m}$ of $M(\xi^\ast, \theta)$ such that the inequality
    \begin{align}
      \label{EquTheoremMED2}
      &\int_{\mathcal{C}(\theta)} \bigl(\varphi((c,d),\xi^\ast,\theta)\bigr)^{q-1}
        \alpha^2\bigl((c_0,d_0), (c,d),\xi^\ast,\theta\bigr)\,d \mu(c,d) - \phi_{\mathrm{MED}_q}^q(\xi^\ast,\theta)\leq 0 \\
      &\text{with  } \alpha\bigl((c_0,d_0),(c,d),\xi,\theta\bigr)= \left(\frac{\partial}{\partial \theta} \eta((c,d),\theta)\right)^T G^T M(\xi,\theta)G\left(\frac{\partial}{\partial \theta} \eta((c_0,d_0),\theta)\right),\nonumber
    \end{align}
    holds for all $(c_0,d_0)\in\mathcal{Z}$.
    Moreover, equality is achieved in \eqref{EquTheoremMED2} for all $(c,d)$ in the support of the design $\xi^*$.
  \end{enumerate}
\end{thm}

\begin{bemerkung}
If the $L_1$-norm is used in \eqref{MED-crit}, the criterion,  $\phi_{\mathrm{MED}_1}$, simplifies to an A-optimality criterion (see e.g. \cite[p.~137]{pukelsheim}). More precisely, $\phi_{\mathrm{MED}_1}$ can be rewritten by
\begin{align*}
    &\int_{\mathcal{C}(\theta)} \left(\frac{\partial}{\partial \theta} \eta((c,d),\theta)\right)^T M^{-}(\xi,\theta)\left(\frac{\partial}{\partial \theta} \eta((c,d),\theta)\right) d\mu(c,d)\\
    =& \int_{\mathcal{C}(\theta)} \Tr\biggl[M^{-}(\xi,\theta) \left(\frac{\partial}{\partial \theta} \eta((c,d),\theta)\right) \left(\frac{\partial}{\partial \theta} \eta((c,d),\theta)\right)^T\biggr]d\mu(c,d)\\
    =& \Tr\biggl[ M^{-}(\xi,\theta) \underbrace{\int_{\mathcal{C}(\theta)} \left(\frac{\partial}{\partial \theta} \eta((c,d),\theta)\right) \left(\frac{\partial}{\partial \theta} \eta((c,d),\theta)\right)^T d\mu(c,d)}_{=: B} \biggr] = \Tr \left[M^{-}(\xi,\theta) B\right] ,
\end{align*}
where the design $\xi$ has to satisfy $\mbox{range}(B) \subset \mbox{range}(M(\xi, \theta))$.
\end{bemerkung}

\begin{bemerkung}
  \label{Remarkefficiency}
  In order to investigate the quality of a (not necessarily optimal) design $\xi$ for the purpose of identifying effective dose combinations, we consider the $\phi_{\mathrm{MED}_q}$-efficiency
  \begin{equation}\label{def:eff}
    \mathrm{eff}_{\phi_{\mathrm{MED}_q}}(\xi) := \frac{\phi_{\mathrm{MED}_q}(\xi^*,\theta)}{\phi_{\mathrm{MED}_q}(\xi,\theta)} \in (0,1] \, .
  \end{equation}
  The quantity $\mathrm{eff}_{\phi_{\mathrm{MED}_q}}$ has an intuitive interpretation.
  Consider an experiment with $N$~observations that will be performed according to the approximate design~$\xi^*$.
  Then $\frac{1}{N}\phi_{\mathrm{MED}_q}(\xi^*,\theta)$ can be interpreted as the approximate value of the mean predictive variance based on the experiment.
  Now, suppose $\xi$ is a different design, for which we could take $M$ observations instead.
  Both approaches will result in the same mean predictive variance if, and only if
  \begin{align*}
    \frac{\phi_{\mathrm{MED}_q}(\xi^\ast,\theta)}{\phi_{\mathrm{MED}_q}(\xi,\theta)}= \frac{N}{M}.
  \end{align*}
  In other words, when~$\xi$ has an efficiency of~$e$,
  we would need to take a factor of~$1/e$ more observations to obtain the same precision as under the locally $\mathrm{MED}_q$-optimal design~$\xi^*$.
\end{bemerkung}

\begin{korollar}
  \label{corollary-elb}
  For an arbitrary design $\xi$ we have
  \begin{align}
    \label{eq:elb}
    \mathrm{eff}_{\phi_{\mathrm{MED}_q}}(\xi) \geq 1- \frac{\underset{(c_0,d_0)\in\mathcal{Z}}{\max} \Psi\bigl(\xi,(c_0,d_0)\bigr)}{\phi_{\mathrm{MED}_q}(\xi)} =: \mathrm{elb}(\xi),
  \end{align}
  where $\Psi(\xi,(c_0,d_0))$ corresponds to the left hand side of inequality~\eqref{EquTheoremMED1} if the information matrix is non-singular,
  or respectively of inequality~\eqref{EquTheoremMED2} in Theorem \ref{equivtheorem} for the singular case.
  Consequently, $\mathrm{elb}(\xi)$ is a lower bound for the efficiency of a design $\xi$ in terms of $\Psi$.
  Notably, the lower bound does not depend on the optimal design $\xi^\ast$.
\end{korollar}
The statement in \eqref{corollary-elb} follows, for example, from rearranging Equation~(2.63) in \cite[p.~67]{fedorov2013optimal}, which is applicable here because the $\mathrm{MED}_q$-criterion in \eqref{MED-crit} is convex and Theorem~\ref{equivtheorem} holds.

\medskip

So far, the optimality criterion for identifying an effective dose combination depends on the unknown, but true parameter $\theta$. Thus, the corresponding locally optimal designs require a-priori information about the parameter vector. In dose combination studies, preliminary knowledge regarding the individual dose-response relationships and the corresponding parameters might be available from earlier dose-finding studies. Moreover, locally optimal designs can be applied as benchmarks for commonly used designs and can serve as starting points for constructing more robust designs with respect to model assumptions. A robust version of the $\phi_{\mathrm{MED}_q}$-criterion can be developed following a Pseudo-Bayesian approach (see e.g. \cite{pronzato1985robust}, \cite{chaloner1989bayesian} and \cite{ChalonerLarntz}). More precisely, let $\pi$ be a prior distribution of the unknown parameter $\theta \in \Theta$, then a design $\xi^\ast$ is called Bayesian $\mathrm{MED}_q$-optimal, if it minimizes the function
\begin{align}
\label{eq:bayescrit}
    \psi(\xi)= \int_\Theta  \phi_{\mathrm{MED}_q}(\xi,\theta) d\pi(\theta)
\end{align}
for all possible designs $\xi\in\Xi$, where $\Xi$ denotes the set of all possible approximate designs on the design space $\mathcal{Z}$.
For the Bayesian version of the $\mathrm{MED}_q$-criterion, analogous versions of the equivalence theorem in Theorem \ref{equivtheorem} and efficiency bounds such as in Corollary \ref{corollary-elb} can be derived.

\section{Case study}
\label{Case study}
In oncological research, test substances are examined to investigate whether they can reduce the growth of tumors. For this reason, the Tumor Growth Inhibition (TGI) [\%] is often used as the primary endpoint.
When $\mathrm{TV}_{t,i,j}$ denotes the tumor volume of animal $j=1,\ldots,n_i$ in treatment group $i=1,\ldots,d$ at day $t=0,\ldots,T$, then $\mathrm{TGI}_{t,i,j}$ is defined as

\begin{align*}
\mathrm{TGI}_{t,i,j} = \left(1- \frac{\mathrm{TV}_{t,i,j}- \mathrm{TV}_{0,i,j}}{\mathrm{med}_j(\mathrm{TV}_{t,0,j})- \mathrm{med}_j(\mathrm{TV}_{0,0,j})}\right)\cdot 100 \,\%.
\end{align*}
Here, $t=0$ refers to measurements at baseline and $i=0$ indicates the negative control and placebo group.

Hence, a TGI value of 100\% denotes a stasis with no observable tumor growth, whereas a TGI value greater than 100\% signifies a reduction (regression) in the tumor volume.
The following example of a mouse experiment consists of two test drugs $C$ and $D$, with $3$ and $4$ dose levels and corresponding negative control observations within the design region $\mathcal{Z}=\left[0,20\right]\times\left[0,7\right]$.
In addition to these data from the monotherapies, some combinations are used and the TGI is measured. The corresponding design, denoted as original design, is displayed in Appendix Table \ref{tab:Designs}.

The data set for this example is derived from 4 pooled experiments of a tumor model, maintaining important parameters such as cell line, strain, and route of administration consistent across all experiments. TGI is evaluated on day 21 and normalized against the control group in each of the experiments. 
By pooling these experiments, we achieve a relatively high total sample size for preclinical experiments with unbalanced sample sizes in the individual dose groups. The pooling was beneficial to maximize available prior knowledge from existing experiments. This allowed for optimal use of the data and provided a solid foundation to justify the framework and specify the ingredients to plan the optimal design.

\begin{figure}
    \centering
    \includegraphics[width=0.5\linewidth]{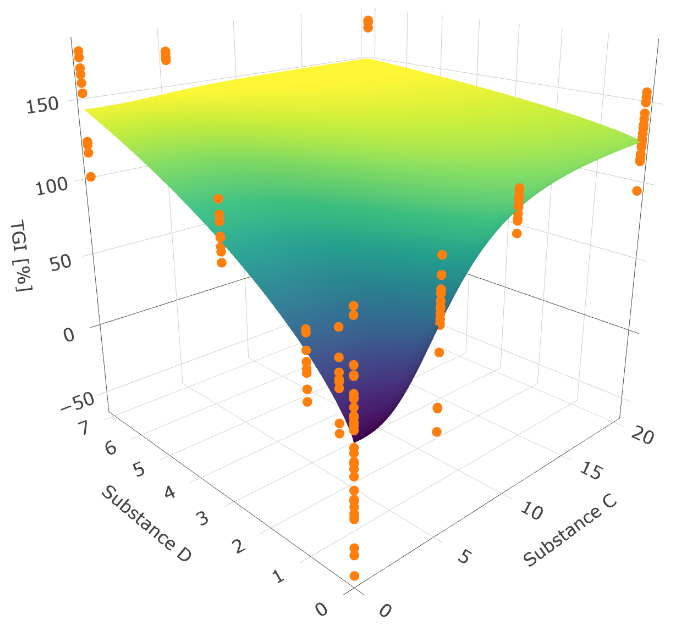}
    \caption{Drug combination surface model for the TGI in \% for case study data with sigmoid Emax model for both substances.
      Additionally, the corresponding data points are displayed in orange.}
    \label{fig:BIData_SigEmaxSigEmax}
\end{figure}

The corresponding data set was utilized to fit a dose combination model with an additive approach as defined in \eqref{interactmod}, resulting in a lower AIC than Greco model defined in \eqref{GrecoMod}. To investigate which dose response models describe the monotherapies best, the multiple comparison and modeling approach from \cite{bretz2005combining} was exclusively applied to the data points representing each monotherapy in a first step. Based on the resulting test statistic, the optimal model selected for both substances was the sigmoid Emax model. Then in a second step, all data points were used to fit a model via maximum likelihood estimation for the corresponding dose combination model, where the corresponding maximum likelihood estimate was determined with
$\theta_0=19.05$, $\mathrm{E}_{\max,C} =111.10$, $\mathrm{ED}_{50,C}=5.83$, $h_C=2.86$, $\mathrm{E}_{\max,D}=410.82$, $\mathrm{ED}_{50,D}=20.00$, $h_D=0.78$, $\gamma=-0.0075$. A visualization of the model and the corresponding data points is given in Figure \ref{fig:BIData_SigEmaxSigEmax}.
Please note that the interaction effect $\gamma$ is slightly negative ($\gamma=-0.0075$), such that the negative interaction would be visible in the plotted response structure only for much higher dose combinations outside the design space $\mathcal{Z}$.
This may be attributed to the predominance of data points from the monotherapies in the case study, with only a limited number of measurements for dose combinations $(c,d)\in \mathcal{Z}$, where the individual doses $c, d$ are not the placebo doses.

\section{Numerical results}\label{sec:Num}

In the following sections, we compare the performance of various design approaches from both theoretical and practical perspectives. The theoretical comparison focuses on efficiencies of the different design approaches, while the practical aspect is explored through simulation studies.

We compare conventional designs, such as factorial and ray designs, against the newly derived $\mathrm{MED}_q$-optimal designs in terms of their precision in estimating effective dose combinations. For the comparison, we restrict ourselves to the $L_2$-norm in the criterion resulting in locally $\mathrm{MED}_2$-optimal designs to which we refer as $\mathrm{MED}$-optimal designs for the sake of brevity. The considered factorial or ray designs motivated from practical aspects are visualized in the Appendix Figure \ref{fig:FactorialRayVisualization}. Moreover, we compare these designs to the corresponding locally D-optimal designs, which are based on the most popular optimality criterion in optimal design theory. The initial comparison is based on the real case study introduced in Section \ref{Case study}, followed by evaluating another scenario in which we consider a different dose combination response relationship given by the Greco model.

To determine $\mathrm{MED}$-optimal designs in practice,
we use the Julia package Kirstine.jl~\citep{SandigKirstine},
where we implement~\eqref{MED-crit} and~\eqref{EquTheoremMED1} via a custom design criterion. Since this package does not support the generalized inverses that occur in~\eqref{EquTheoremMED2}, we limit our investigation to scenarios where~\eqref{EquTheoremMED1} is applicable. Hence the resulting optimal designs always allow for the estimation of the full unknown model parameter.
For the measure~$\mu$, we choose a uniform discrete distribution on a finite number of points in~$\mathcal{C}(\theta)$,
which we obtain by first evaluating the response function on a grid,
and then interpolating the contour lines with the marching squares algorithm~\citep{lorensen-1987-march}.
To verify that a numerically obtained designs is (almost) optimal,
we compute the efficiency lower bound~\eqref{eq:elb} obtained in Corollary \ref{corollary-elb} via the maximum of~$\Psi$ over a $101\times101$-point grid on~$\mathcal{Z}$.
For all considered $\mathrm{MED}$-optimal designs, the corresponding efficiency bound is at least~$0.99$
(see the tabulated designs in the Appendix).
The subsequent simulations and analyses were performed in R~\citep{RCoreTeam2024}.
Our code and the computed designs can be found in Zenodo at~\citep{schuermeyer-2025-code-optim}.

\subsection{Efficiency-based comparison for the case study data}

It is well established that the selection of dose combinations has a direct impact on the precision of estimation. However, it is crucial to investigate the extent of this influence. The quality of different designs in terms of their ability to estimate the effective dose combinations can be theoretically assessed based on their associated $\phi_{\text{MED}}$-efficiency in \eqref{def:eff} (see Remark~\ref{Remarkefficiency} for details). 
In this context, a higher $\phi_{\text{MED}}$-efficiency indicates a superior design with respect to the corresponding $\mathrm{MED}$-criterion.

In the following, the performance of the different design approaches is compared theoretically based on their corresponding $\phi_{\text{MED}}$-efficiencies for the parameter estimates of the case study data if two sigmoid Emax models are used (see Section \ref{Case study} for details). In this context, a precise estimation of the effective dose combinations leading to 10\% and 50\% of the maximal effect in terms of TGI is of interest, to investigate the initital TGI and a bisection of the TGI in terms of the maximal effect within the design space. Furthermore, a second set of effective concentrations is of interest to define the magnitude of the maximal effect of the TGI, i.e. where 20\% and 80\% of the maximal effect are achieved.

\begin{figure}
    \centering
    \subfigure[Locally $\mathrm{MED}$-optimal design with $\mathrm{MED}_{10}$ and $\mathrm{MED}_{50}$ sets.]{\includegraphics[width=0.49\textwidth]{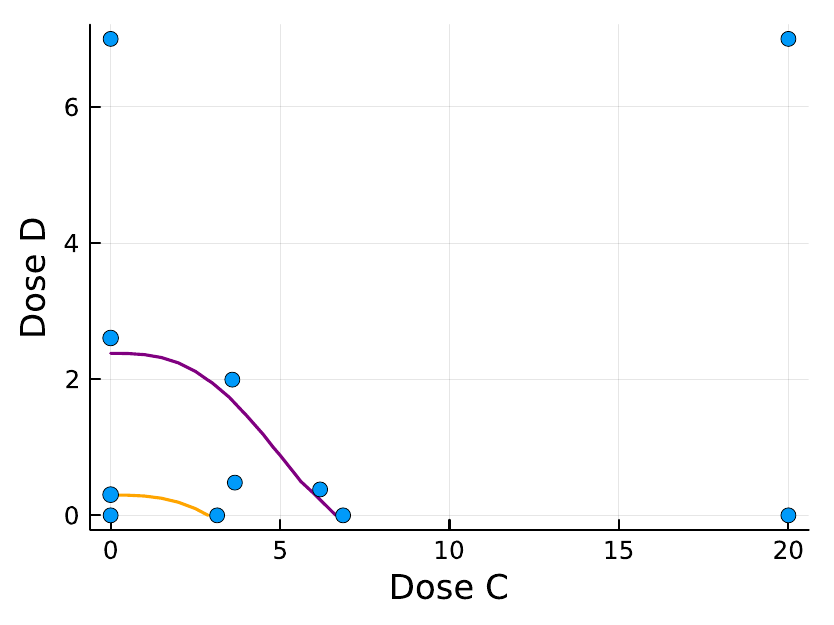}}
    \subfigure[Locally $\mathrm{MED}$-optimal design with $\mathrm{MED}_{20}$ and $\mathrm{MED}_{80}$ sets.]{\includegraphics[width=0.49\textwidth]{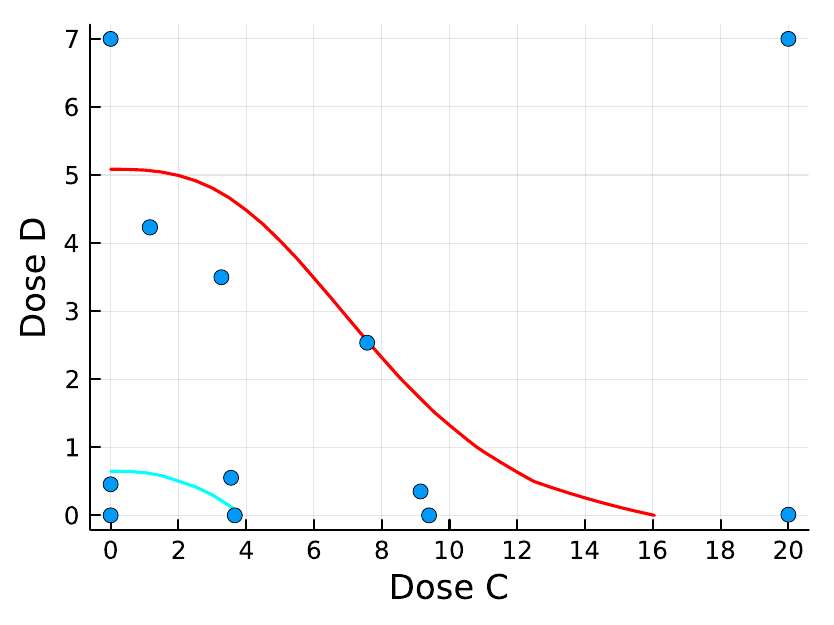}}
    \subfigure[D-optimal design with $\mathrm{MED}_{10}$ and $\mathrm{MED}_{50}$ sets.]{\includegraphics[width=0.49\textwidth]{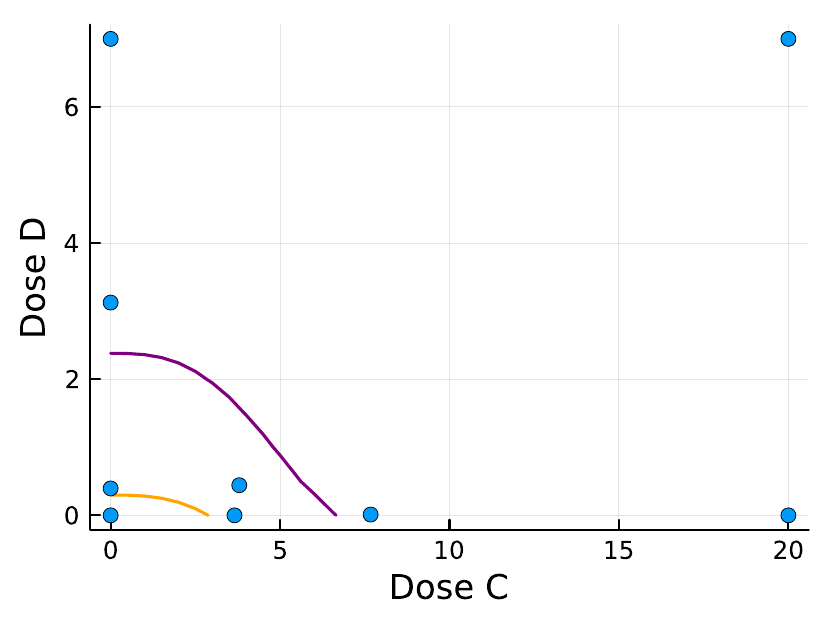}}
    \subfigure[D-optimal design with $\mathrm{MED}_{20}$ and $\mathrm{MED}_{80}$ sets.]{\includegraphics[width=0.49\textwidth]{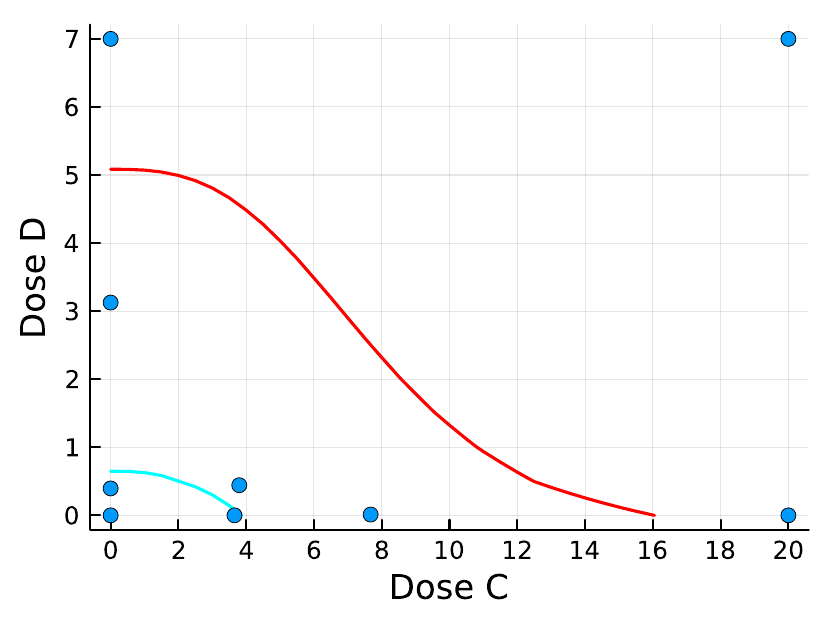}}
    \caption{Visualization of the considered locally MED-optimal designs and D-optimal design for the case study, which includes their support points represented by dots, with their sizes corresponding to their weights.}
    \label{fig: VisualizationDesignsCaseStudy}
\end{figure}

For the comparison, we consider the locally $\mathrm{MED}$-optimal design for the effective doses $\text{MED}_{10}$ and $\text{MED}_{50}$ using the parameter estimate of the case study and conventional designs as the factorial $4\times 4$ design or the ray designs. 
Furthermore, we compare the designs to the corresponding locally D-optimal design. 
Note that we do not consider the $3\times3$ factorial design in the situation of the case study, as the parameter of the considered model cannot be estimated uniquely.
Moreover, the considered locally $\mathrm{MED}$-optimal designs and the D-optimal design are illustrated in Figure \ref{fig: VisualizationDesignsCaseStudy}, along with the corresponding contour lines of interest. The locally $\mathrm{MED}$-optimal design results in dose combinations with higher weights close to the contour lines of interest, whereas the D-optimal design, by definition, aims to capture the overall response relationship.

The efficiencies of all considered designs, evaluated according to the multivariate effective dose criterion on different sets of contour lines, are presented in Table \ref{Tab:EfficiencyComparison}. By definition, the highest efficiencies across the different criteria are achieved by their corresponding optimal designs.
For a precise estimation of effective doses at the 10\% and 50\% levels, the D-optimal design demonstrates the next highest efficiency, followed by the locally $\mathrm{MED}$-optimal design for the 20\% and 80\% contour levels, and then the original design. However, all three designs exhibit efficiencies that are at most 60\%, indicating an insufficient performance compared to the optimal one. Furthermore, an efficiency of at most 60\% for all other designs demonstrates that theoretically more than 65\% more observations are necessary for all considered designs to achieve the same precision as the locally $\mathrm{MED} (10,50)$-optimal design. The factorial design and both ray designs show very low efficiencies, demonstrating inadequate capability to estimate those contour lines and thus determine the corresponding effective concentrations accurately.

\begin{table}
    \centering
    \caption{$\phi_{\mathrm{MED}_2}$-efficiencies of the considered designs for different effect levels based on the case study.}
    \label{Tab:EfficiencyComparison}
    \begin{tabular}{lcc}
      \toprule
      \multirow{2}{*}{Design} & \multicolumn{2}{c}{Levels} \\
      \cmidrule{2-3}
             & $(10, 50)$ & $(20, 80)$ \\

      \midrule
      Ray $4/2$ & $0.07$ & $0.09$ \\
      Ray $4/4$ & $0.06$ & $0.08$ \\
      Factorial $4\times 4$ & $0.04$ & $0.06$ \\
      Original & $0.34$ & $0.47$ \\
      D-optimal & $0.6$ & $0.8$ \\
      MED $(10, 50)$ & $1.00$ & $0.19$ \\
      MED $(20, 80)$ & $0.52$ & $1.00$ \\
      \bottomrule
    \end{tabular}
\end{table}

For the $\mathrm{MED}$-criterion on 20\% and 80\% contour lines, similar results are obtained. Notably, the D-optimal design exhibits an efficiency of 80\%, indicating good performance in this setting. Additionally, the original design shows a higher efficiency than in the first setting, indicating a better ability to identify these contour lines. However, its efficiency remains below 50\%, indicating an insufficient performance. For the original design, twice as many observations are required to achieve theoretically the same precision as the corresponding locally $\mathrm{MED}$-optimal design. The locally $\mathrm{MED}$-optimal design that aims at estimating the first set of contour lines (i.e. 10\% and 50\%) most precisely, exhibits an efficiency below 20\%, demonstrating poor performance for the second set of contour lines. This indicates that a misspecified design can perform poorly for contour lines or other settings for which it was not initially designed. Note that therefore the robustness of locally $\mathrm{MED}$-optimal designs is considered in Section \ref{robustnessAnalysis}. However, the commonly used factorial and ray designs perform even worse in the analysis of the case study.

\subsection{Simulation study}
\label{SectionSimulationStudy}

We perform a simulation study to compare the different design approaches for different types of interaction and different dose-combination response relationships. In particular, we investigate how precisely the contour lines of the effective dose combinations can be estimated for the different design approaches with existing variability of the data points.

\begin{figure}
    \centering
    \subfigure[Scenario~1]{\includegraphics[width=0.32\textwidth]{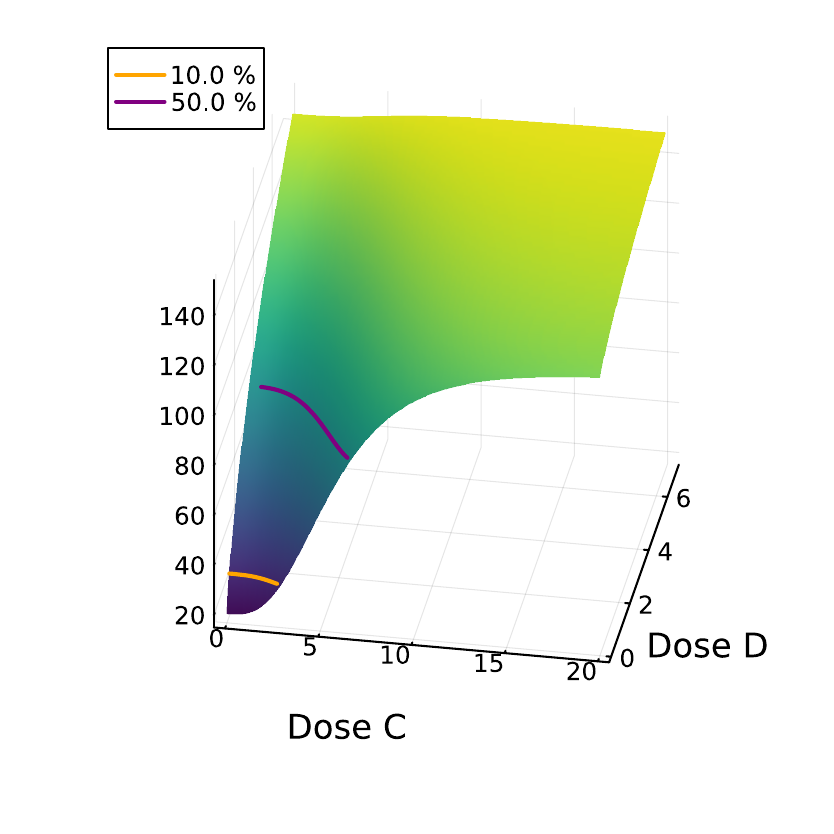}}
    \subfigure[Scenario~2]{\includegraphics[width=0.32\textwidth]{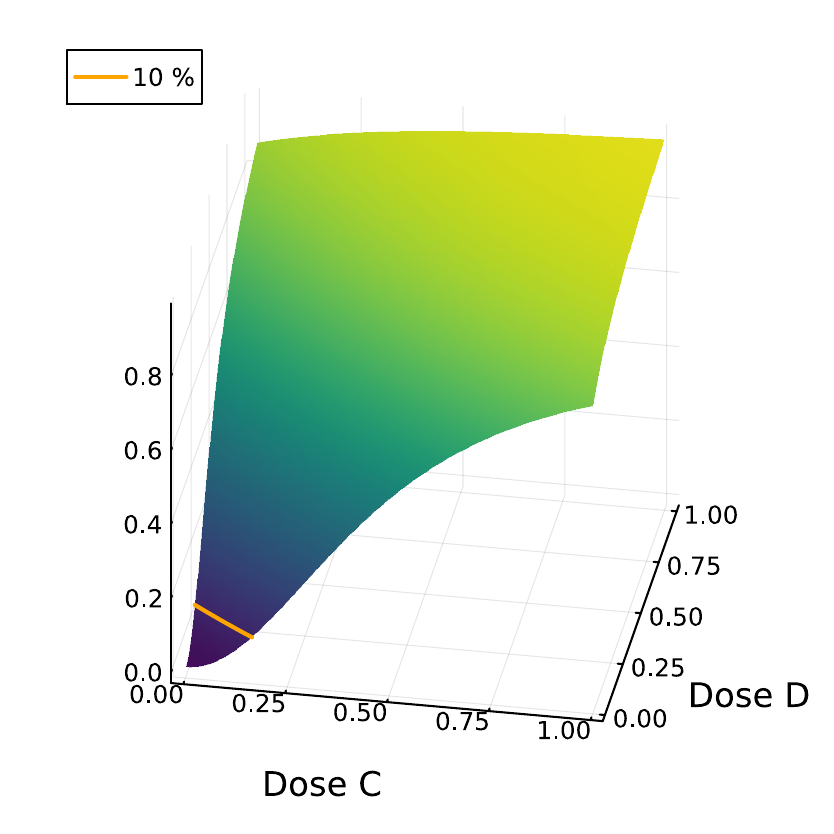}}
    \caption{Response-surface models for the two considered scenarios with considered contour lines.}
    \label{fig: SUrfaceScenarios}
\end{figure}

The considered parameters and designs of two scenarios are shown in Table \ref{tab:SimulationScenarios}. Additionally, the scenario-specific response surface relationships are visualized in Figure \ref{fig: SUrfaceScenarios}. The individual scenarios exhibit distinct differences in their response surfaces. Scenario~1 is motivated by the case study (see Section \ref{Case study}) and shows no visible interaction of the two considered substances within the design space due to the slightly negative interaction effect of~$\gamma=-0.0075$. Here, the contour levels of 10\% and 50\% can be achieved with doses from each monotherapy. Conversely, Scenario~2 demonstrates a clear positive interaction. In this scenario, the initial onset of the combined effect of both substances, resulting in the 10\% contour line, is of interest. While the Scenario~1 is modeled by the additive approach in \eqref{interactmod}, in Scenario~2  the Greco model in \eqref{GrecoMod} is investigated.

\begin{table}
  \centering
  \caption{Parameter specification of different simulation scenarios.}
  \label{tab:SimulationScenarios}
  \begin{tabular}{lll}
    \toprule
                                        & Scenario~1                & Scenario~2              \\
    \midrule
    Model basis                         & Case study                & Example Choice             \\
    Assumed model                       & Additive model                & Greco model             \\
                                 & (sigmoid Emax,              &                          \\
                                & sigmoid Emax)              &                                 \\
    Design space                        & $[0,20]\times[0,7]$       & $[0,1]\times[0,1]$  \\
    Parameter                      & $\theta_0=19.05$                   & $\theta_0=0$, $\mathrm{E}_{\max}=1, \gamma = 2$                                  \\
                         & $\theta_C=(111.10, 5.83, 2.86)$    & $\mathrm{EC}_{50\, C}= 0.5 $                 \\
                         & $\theta_D=(410.82, 20.00, 0.78)$   &                $\mathrm{EC}_{50\, D}= 0.4$    \\
    Interaction                       & $\gamma= -0.0075$                 & $\alpha = 0.2$                         \\
    Contour level(s)         & $(10, 50)$                & $10$                  \\
    Error sd                            & $\sigma_{\mathrm{CS}}=24$ & $\sigma=0.45$                   \\
    \midrule
    Factorial $3\times 3$               & --                        & \checkmark                            \\
    Factorial $4\times 4$               & \checkmark                & \checkmark                      \\
    Ray 3/2                             & --                        & \checkmark                             \\
    Ray 4/2                             & \checkmark                & \checkmark                              \\
    Ray 4/4                             & \checkmark                & --                            \\
    D-optimal                           & \checkmark                & \checkmark                      \\
    $\mathrm{MED}$-optimal              & \checkmark                & \checkmark                      \\
    Misspecified $\mathrm{MED}$-optimal & \checkmark                & --                                     \\
    \bottomrule
  \end{tabular}
\end{table}

We consider different total sample sizes $N\in \{27,36, 45, 90, 152\}$ for both scenarios. For some designs, e.g., the factorial $3\times 3$ design, this leads to an equal number of observations at each dose combination for every choice of $N$. For all the remaining designs, an efficient rounding procedure of \cite{pukelsheimrieder1992} is used to obtain integer numbers at each dose combination. For both scenarios under consideration, $s=1000$ simulation iterations are performed, where the different design approaches are compared in terms of their precision.\\

The general structure of the simulation setup remains the same for both scenarios and consists of the following steps:\\
Initially, the model and parameter settings are specified for each scenario, respectively (see Table \ref{tab:SimulationScenarios}). In detail, this corresponds to the assumed combination model $\eta((c,d),\theta_t)$ and the parameter setting $\theta_t$. 
For each simulation truth $N$ new data points with normally distributed errors $\varepsilon\sim \mathcal{N}(0,\sigma^2)$ of the given variance $\sigma^2$ are sampled at the dose combination levels of the considered design.

Then, the same dose combination model is fitted to the new data points in simulation step $s$, resulting in the model fit $\eta((c,d),\hat\theta_s)$. For this model the $k$ effective dose combinations sets $\mathrm{MED}_{p_i}$ are calculated for the scenario's levels $p_i$, $i=1,\ldots,k$. In the next step, the Root Mean Squared Error (short: RMSE) is calculated for all contour lines of interest via
\begin{align*}
  &\mathrm{RMSE} = \frac{1}{k}\sum_{i=1}^k\sqrt{\frac{1}{l_i} \sum_{j=1}^{l_i}  \bigl(p_i-\eta((c_j,d_j)_i,\theta_t)\bigr)^2} \\
  & \text{with } (c_j,d_j)_i\in \mathrm{MED}_{p_i}(\hat\theta_s) \quad  \text{for } i=1,\ldots, k \text{ and } j=1,\ldots,l_i\,.
\end{align*}
where $l_i$ denotes the length of the set $\text{MED}_{p_i}(\hat\theta_s)$ that is calculated based on a grid, $i=1, \ldots, k$.
By means of this, the precision of the estimated effects at the MED sets is measured for each simulation step in comparison to the initial ``true'' model.

It is important to note that due to the complex model setup, computational issues may arise when fitting the model or computing the MED sets for the new simulated data points during the simulation step. These cases were excluded from further analysis. If such cases were identified in the analysis, they will be stated.\\

\begin{figure}
    \centering
    \includegraphics[width=0.8\linewidth]{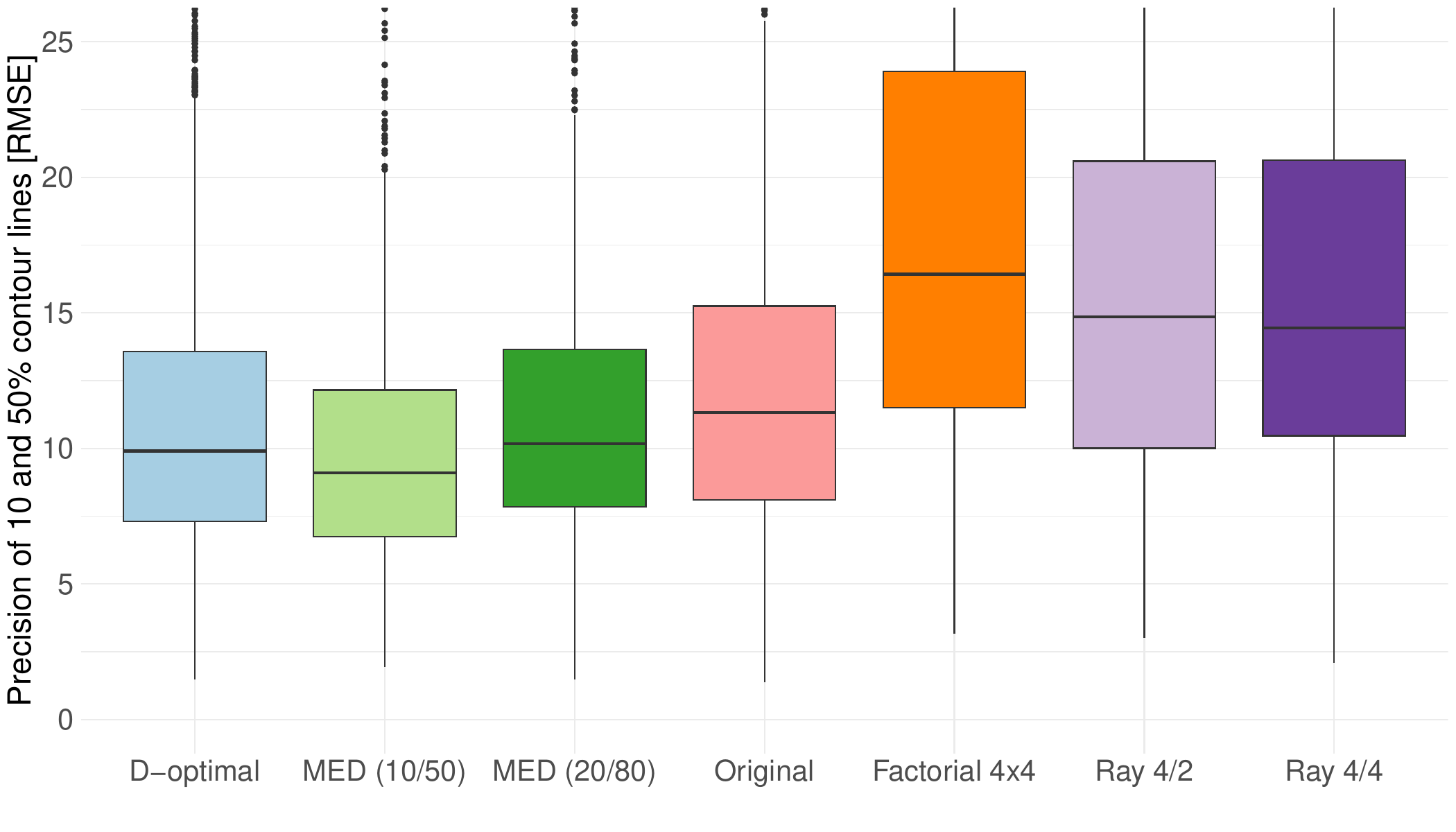}
    \caption{Extract of simulation RMSE values in Scenario~1 (case study) regarding the 10\% and 50\% contour levels for $N=27$ measurements.
    The locally $\mathrm{MED} (10,50)$-optimal design outperforms all other designs, especially the traditionally used factorial design and both ray designs. The locally $\mathrm{MED} (10,50)$-optimal design shows the smallest RMSE-values and therefore the highest precision for prediction at the set of effective dose combinations. Additionally, the smallest variability can be seen by the locally $\mathrm{MED} (10,50)$-optimal design.
    }
    \label{fig:SImResults_BIData27}
\end{figure}

Analyzing Scenario~1 reveals how well the different designs perform in the practically motivated setting of the case study (see Section \ref{Case study} again). Since the case study serves as the source for this scenario, the initial parameter setting corresponds to the model fit of the case study. The corresponding results for the simulation study of Scenario~1 can be seen exemplarily for 27 measurements in Figure \ref{fig:SImResults_BIData27}, where the precision of estimating the 10\% and 50\% contour lines is shown via the corresponding RMSE values.

Both locally $\mathrm{MED}$-optimal designs demonstrate strong performance, particularly when compared to the traditional factorial or the ray designs. The locally $\mathrm{MED}$-optimal design, specifically tailored for estimating the corresponding 10\% and 50\% contour lines, exhibits the smallest median RMSE value and variability, outperforming all other design approaches. The smaller the RMSE values are, the more precisely the contour lines can be estimated. This is also directly linked to a precise estimation of the corresponding MED sets. Hence, the considered $\mathrm{MED}$-optimal design leads to the highest precision of the effective concentrations. Furthermore, the locally D-optimal design shows the next best performance. The locally $\mathrm{MED}$-optimal design not intended for estimating the contour lines of interest at 10\% and 50\%, but rather
for the 20\% and 80\% contour lines 
also performs well and outperforms the factorial design and the ray designs. 
Overall, the factorial design and the ray designs are clearly inferior to the other design approaches in this context, as indicated by their higher RMSE values and variability. However, it is important to note that both ray designs achieve higher precision than the factorial $4\times4$ design, despite utilizing fewer dose combination levels.

A similar pattern is observed for the other sample sizes. The locally $\mathrm{MED}$-optimal design for estimating the 10\% and 50\% contour line outperforms the other designs for all numbers of observations. As the number of measurements increases, both the RMSE values and the variability of all designs decrease, indicating higher precision for all designs (see Appendix Figure \ref{fig:SImResults_BIData}). Nevertheless, the factorial or ray designs remain inferior to the other design approaches. The misspecified locally $\mathrm{MED}$-optimal design for the contour lines of 20\% and 80\% also outperforms the original design for higher sample sizes in terms of the median RMSE values and ranks third best in this case. Additionally, it is important to note that the factorial $4\times 4$ design and both ray designs result in very large outliers, particularly for small sample sizes.

Furthermore, it should be noted that issues may arise when fitting the models during the simulation due to the model's complexity. Across all $N$ and all replications, one model could not be estimated for all simulations if the original design, or a ray design was used. Additionally, four problematic simulation steps were observed in the factorial design. Given that 1000 simulation steps are performed for each different sample size and only a few problematic steps were encountered in total, these can be considered negligible. 
Both locally $\mathrm{MED}$-optimal designs and the locally D-optimal design did not exhibit any problems when estimating the models during the simulation steps.
\medskip


In the Scenario~2 a Greco model with a clear positive interaction effect is considered (see Figure \ref{fig: SUrfaceScenarios}). Here, we consider the situation where the precise estimation of one contour line, i.e. 10\%, is of interest.
Similar to Scenario~1, the different design approaches (see Table \ref{tab:SimulationScenarios} and Appendix Table \ref{tab:DesignsScenario2}) are compared. Additionally, the locally $\mathrm{MED}$-optimal and D-optimal designs are displayed in Appendix Figure \ref{fig: VisualizationDesignsScenario2}.

\begin{figure}
    \centering
    \includegraphics[width=0.8\linewidth]{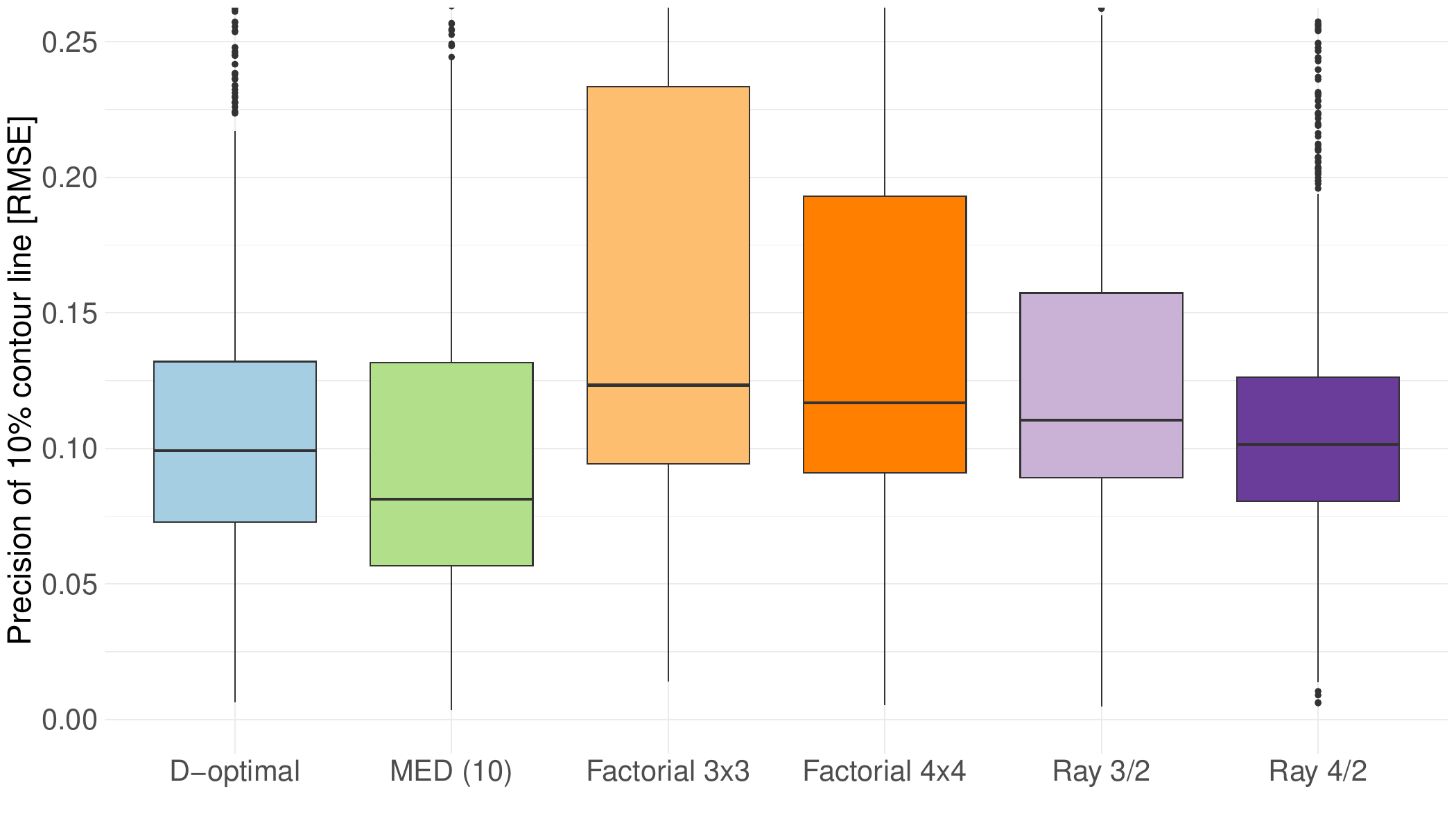}
    \caption{Extract of simulation RMSE values in Scenario~2 (Greco model) regarding the 10\% contour level for $N=27$ measurements.
    The locally $\mathrm{MED}$-optimal design outperforms all other designs. It shows the smallest RMSE-values and therefore the highest precision of the set of effective dose combinations.
    }
    \label{fig:SImResults_Scenario527}
\end{figure}

The precision of the considered design approaches regarding the contour line of interest in Scenario~2 is shown in Figure \ref{fig:SImResults_Scenario527} for the case of 27 measurements. Clearly, the locally $\mathrm{MED}$-optimal design outperforms all other designs in terms of the lowest RMSE values, indicating the highest precision. The next highest precision is achieved by the D-optimal and the Ray 4/2 design. Both designs show a similar performance; however, the RMSE values of the Ray 4/2 design vary less than those of the D-optimal design. Furthermore, the Ray 3/2 design ranks fourth in Scenario~2, but is considerably less effective than the locally $\mathrm{MED}$-optimal design. Both factorial designs exhibit much higher RMSE values and greater variability than all other designs, indicating that they are not suitable for estimating $\mathrm{MED}$ curves if the $10\%$ contour line is used in combination with the Greco model.

While precision generally improves with an increasing number of measurements used in the simulation, the factorial and ray designs still perform poorly relative to the theoretically motivated approaches based on design theory, i.e. the D-optimal and the $\mathrm{MED}$-optimal design (see Appendix Figure \ref{fig:SImResults_Scenario2ALL}). All practically motivated designs show RMSE values twice as high as those of the locally $\mathrm{MED}$-optimal design.

In this scenario, issues regarding the model fit were encountered for any of the design choices in the simulation study. The Factorial $4\times 4$, the locally $\mathrm{MED}$-optimal and the D-optimal design exhibited 1, 2, 3 problematic simulations, respectively, across all sample sizes $N$. The Factorial $3\times 3$ design showed 9 problematic simulations. Both ray designs yielded the highest number of problematic simulation steps, which are 16 for the Ray 4/2 design and 64 for the Ray 3/2 design. Although these numbers are negligible relative to the total of 1000 simulation steps, they nonetheless indicate that the ray designs may be less suitable compared to the other design strategies.

In summary, both scenarios demonstrate that using the $\mathrm{MED}$-optimal design allows for more precise estimation of effective concentrations compared to the other designs, even though all designs employ the same sample size. Consequently, the alternative designs would require a larger number of observations (animals) to achieve the same level of precision in estimating effective concentrations as obtained with the $\mathrm{MED}$-optimal design. Therefore, the $\mathrm{MED}$-optimal design should be preferred over traditionally used designs like Factorial or Ray designs.

\subsection{Robustness analysis}
\label{robustnessAnalysis}

In dose combination experiments, the dose-response relationships of the individual substances are often well understood from previous experiments, but the interaction effect is unknown. Thus, it is reasonable to account for different possible values of the interaction effect when planning the drug combination experiments. To ensure robustness, Bayesian designs can be used to account for uncertainty about the interaction effect. In this section, the performance of the locally $\mathrm{MED}$-optimal design itself and its corresponding Bayesian version defined in \eqref{eq:bayescrit} is investigated in terms of their ability to estimate effective dose combinations most precisely.

First, we investigate a setting where due to previous experiments of each substance exclusively, the dose-response relationship of the monotherapies is already known. Furthermore, the interaction effect is not precisely known but is based on a small preliminary experiment of the combined substances, assumed to be non-negative. We construct a locally $\mathrm{MED}$-optimal design for the simulation truth $\gamma=0.02$ and a locally $\mathrm{MED}$-optimal design for the misspecified value $\gamma=0.05$ to investigate the influence of a missspecified interaction parameter on the estimation precision. Moreover, the Bayesian $\mathrm{MED}$-optimal design for different non-negative values of $\gamma$ is calculated using the definition in \eqref{eq:bayescrit}.

In detail, we investigate an additive model with two Emax models and the assumed parameter setting $\theta_0=0, \theta_C=(80,3), \theta_D=(120,10)$ and a positive interaction effect $\gamma=0.02$. Here, the considered design approaches are evaluated in terms of their ability to estimate the 10\% and 30\% contour lines most precisely. The corresponding surface model and its corresponding contour lines is shown in the Appendix Figure \ref{fig:BayesianSetting1_OldSzenario2}.

To imitate the case of not knowing the real interaction effect before conducting a dose combination experiment, but assuming the interaction effect to be greater or equal to zero, we consider
three different values, i.e. $\{0, 0.01,0.02\}$, all equally weighted for the interaction effect $\gamma$, which for example can be assessed via preliminary experiments or literature research.
The corresponding Bayesian $\mathrm{MED}$-optimal design is displayed in Appendix Figure \ref{fig: VisualizationBayesianDesigns9095} and Table \ref{tab:DesignsScenarioBayesian9095and1030}. Additionally, the resulting contour lines for the different values of $\gamma$ are shown in Appendix Figure \ref{fig:ContourlinesBayesianDesign9095}. Although the response-surface is substantially different for the different choices of the interaction effect, the contour lines in this case look quite similar. Due to this fact, the Bayesian $\mathrm{MED}$-optimal design has very similar support points to both locally $\mathrm{MED}$-optimal designs in this case.

\begin{figure}
    \centering
    \includegraphics[width=0.8\linewidth]{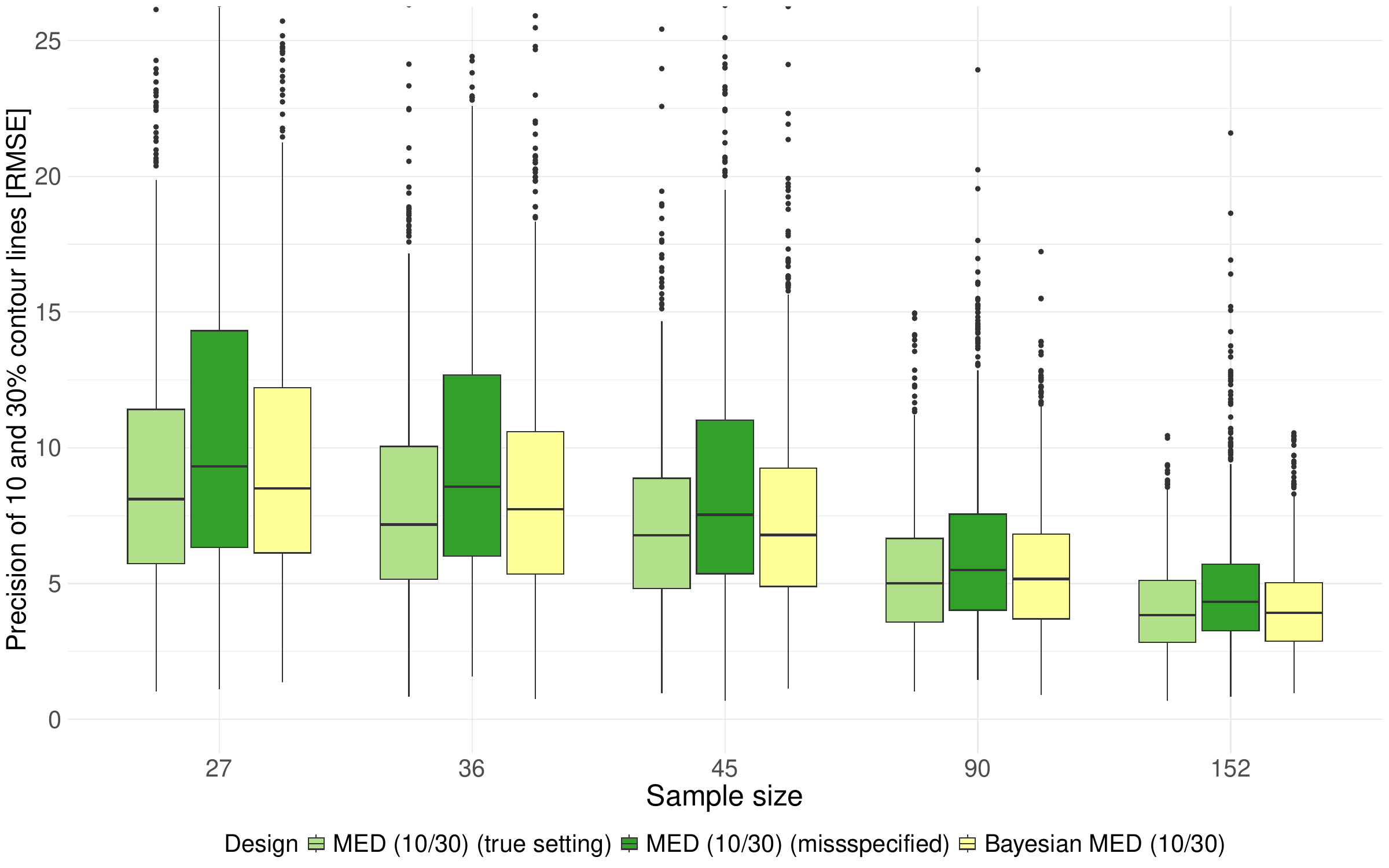}
    \caption{Extract of RMSE values for the robustness analysis setting~1 with simulation truth $\gamma=0.02$ (two Emax models) grouped by design and total sample size for the robustness analysis of MED designs. All designs show a similar ability to estimate the contour lines for $p_1=10$ and $p_2=30$ in this setting.
    }
    \label{fig:SImResults_Scenario2_27_plusBayes}
\end{figure}

In a simulation study, we compare the three different designs (Appendix Table \ref{tab:DesignsScenarioBayesian9095and1030}) in terms of their ability to precisely estimate the 10\% and 30\% contour lines. Figure \ref{fig:SImResults_Scenario2_27_plusBayes} shows the RMSE values of the 10\% and 30\% curves for different sample sizes. All designs seem to perform quite similarly, as the median RMSE values as well as their corresponding variabilities illustrated by the corresponding boxes are similar across the designs. The misspecified locally $\text{MED}$-optimal design, which assumed an interaction effect of $\gamma=0.005$ instead of $\gamma = 0.02$ performs slightly worse for small sample sizes, but shows a similar performance to the other approaches in the case of higher sample sizes. This might be due to the fact that the MEDs also look similar in this setting for different values of $\gamma$, although the corresponding surfaces differ.

Besides, there were no numerical difficulties to estimate the correct specified or Bayesian $\mathrm{MED}$-optimal design, respectively. The missspecified locally $\mathrm{MED}$-optimal design instead showed 3 problematic iteration steps across all samples sizes.\\

In a second setting, we explore the capabilities of the Bayesian MED design when the interaction effect remains uncertain and a broader range of values for $\gamma$ is considered. We consider five different potential values for the interaction effect $\gamma$, namely $\{-0.02,0.01,0,0.01,0.02\}$.

In this setting,  we assume that the true (but unknown before) dose combination relationship exhibits a negative interaction effect with $\gamma_t=-0.01$. The corresponding contour lines of interest are displayed in Appendix Figure \ref{fig:ContourlinesBayesianDesign_prior5} and exhibit widely different shapes. A locally $\mathrm{MED}$-optimal design is calculated for the simulated truth of $\gamma_t=-0.01$ and a misspecified $\gamma=0.02$ as well as the Bayesian $\mathrm{MED}$ design for the considered prior. The designs are displayed in Appendix Figure \ref{fig: VisualizationBayesianDesigns_prior5} and Table \ref{tab:DesignsScenarioBayesianprior5}. 

\begin{figure}
    \centering
    \includegraphics[width=0.8\linewidth]{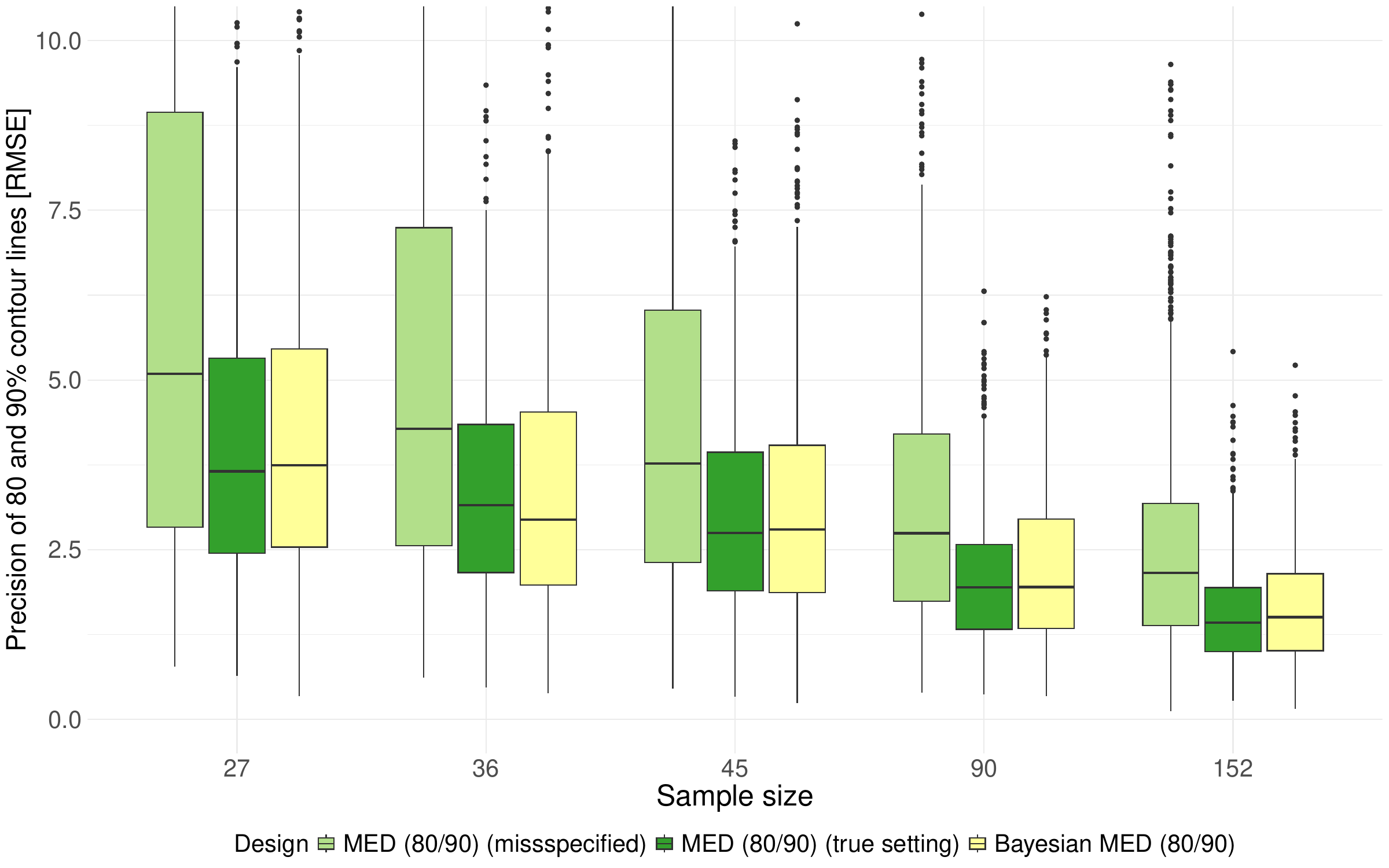}
    \caption{Extract of RMSE values for the robustness setting 2 with simulation truth $\gamma=-0.01$ (two Emax models) grouped by design and total sample size for the robustness analysis of MED designs. The locally MED-optimal design and the Bayesian MED design show a similar ability to estimate the contour lines for $p_1=80$ and $p_2=90$ in this setting, while the misspecified design shows clearly higher RMSE values.
    }
    \label{fig:SImResults_Scenario2_27_plusBayes_prior5}
\end{figure}

Figure \ref{fig:SImResults_Scenario2_27_plusBayes_prior5} illustrates the precision of the 80\% and 90\% contour lines for the second setting in this robustness analysis. In this case, the locally  MED-optimal design, which was tailored to the true parameter setting, performs best as expected and can be treated as the gold standard. The misspecified locally MED-optimal design exhibits considerably higher RMSE values, indicating a lower ability to estimate the contour lines accurately. Conversely, the Bayesian MED-optimal design, which accounts for a range of different values of the interaction effect $\gamma$, demonstrates excellent performance in this context, nearly on par with the locally optimal design.

Besides, there were no difficulties to estimate the Bayesian $\mathrm{MED}$ or both locally $\mathrm{MED}$-optimal designs.

\section{Discussion and conclusion}
Identifying effective dose combinations that achieve a prespecified percentage of the maximal effect has recently become more relevant in drug-combination studies. However, so far, the optimal design of drug-combination experiments is based on the established classical D-optimality criterion (see \cite{papathanasiou2019optimizing}) or criteria related to therapeutic indices, such as the drug combination index (see \cite{hollandletz2018}). The disadvantage of these designs is that they may result in an imprecise estimation of effective dose combinations, as given by contour lines, if nonlinear surface models are used to describe the dose combination response relationship.
Therefore, we propose a novel design approach for drug-combination experiments that directly addresses the precise estimation of effective dose combinations. Here, $\mathrm{MED}_p$ denotes the set of dose combinations resulting in $p\%$ of the maximum effect, i.e. the contour line of a nonlinear surface model at $p\%$. We define a design as optimal for identifying effective dose combinations if it minimizes the $L_q$-norm of the variances of the confidence band describing the effect at the dose combinations contained in $\mathrm{MED}_p$.

\noindent
For this novel criterion, we provide optimal design theory via equivalence theorems and some further analytic results.\\
\noindent
The $\mathrm{MED}_p$-optimal design criterion is a flexible approach, which can be used for a broad class of linear and nonlinear response surface models. Therefore, we investigate in an extensive simulation study the ability of the $\mathrm{MED}_p$-optimal design to estimate contour lines of interest precisely for commonly used response surface models in dose combination studies. In two different simulation scenarios, we explore the performance of the $\mathrm{MED}_p$-optimal design for different response surface models and assumed interactions between the two considered drugs. Moreover, we compare the performance of the $\mathrm{MED}$-optimal design with commonly used designs, including factorial and ray designs. The analysis also incorporates D-optimal designs. Also, one of the simulation scenarios is based on a real case study. Here, a data set of four pooled mouse experiments was used to investigate the dose-response relationship of two oncological substances.

In both scenarios, the optimal designs for identifying effective dose combinations demonstrate strong performance, outperforming the other designs considerably. Our results clearly demonstrate that the $\mathrm{MED}$-optimal design should be preferred over traditionally used designs, as it provides higher precision in estimating the contour lines while using the same number of measurements. Other designs would require additional observations (e.g. animals) to achieve the same level of precision as the $\mathrm{MED}$-optimal design. In research and animal experiments, such as oncology research, this approach can provide notable benefits. Dose-combination studies often involve a considerable number of animals. With an optimized study design, however, it may be possible to achieve precise estimates with smaller sample sizes. This could offer advantages from an ethical perspective by supporting the principles of the 3R framework (Replace, Reduce, Refine) and may also help facilitate the implementation of such experiments.

Also, the locally D-optimal design leads to effectively estimating the contour lines. 
However, it struggles if the scope is the identification of effective dose combinations for one marginal effect level $p$. The D-optimal design aims at achieving a precise estimation of the parameter of interest. Therefore, it appears that when more contour lines are relevant and these contours are more distinct, the D-optimal design is also a suitable choice, as it effectively captures the overall surface structure. 
Furthermore, the practically motivated designs like the factorial or ray designs are clearly inferior to the theoretical motivated designs.

Given that prior knowledge of single-substance dose-response relationships often exists before conducting dose combination studies, while the interaction effects are unknown, we propose a robust version of the developed criterion using a (pseudo) Bayesian approach.  More precisely, instead of assuming a specific value for the interaction effect, several possible values for the interaction effect are incorporated via a prior distribution. Performance comparisons in simulation studies reveal that the Bayesian approach enhances robustness, particularly if the prior knowledge about the interaction effect is less informative. In this case, the contour lines of the surface model vary considerably, and a potentially misspecified locally optimal design for identifying effective dose combinations is inferior to the corresponding Bayesian optimal design.  However, if the prior distribution is informative, reflecting a specific direction of the type of interaction, the differences between the locally optimal and the Bayesian optimal design are minor. Here, the contour lines of interest show slight variation with different values of the interaction effect $\gamma$, and both the locally optimal designs (under slight misspecification) and the corresponding Bayesian design perform similarly. 
In situations where the effect is uncertain before conducting the experiment, and with varying contour lines depending on the interaction effect $\gamma$, the Bayesian optimal designs should be preferred over a potentially misspecified version of the locally optimal design for identifying effective dose combinations.

We emphasize that, even in the one-dimensional case, the confidence interval for the effective doses can be quite large in practical settings. Due to various uncontrollable factors in the laboratory, the effective doses can vary, even between two experiments with similar setups \citep{jiang2014summarizing}. This underscores the inherent variability of the ED itself. This issue can also arise in two-dimensional settings and is a persistent challenge. However, the developed criterion focuses on minimizing the asymptotic variance of the effective dose combinations, which also results in the minimal variability of the estimation of the effective doses.

The proposed design approach is very flexible, allowing to adapt to any nonlinear surface model, including non-monotonic regression functions for the individual drugs under consideration. We focused on identifying multivariate effective dose combinations that lead to  $p\%$ of the maximal effect. However, the approach can be easily be extended to any contour line of interest. To facilitate the application of the developed optimality criterion by practitioners, code examples are provided in Zenodo at~\citep{schuermeyer-2025-code-optim}.

The criterion is based on an $L_q$-norm where $q\in[1,\infty)$ has to be fixed in advance. We analyzed the performance of the optimal designs for various selected values of $q$, finding that, for example, $q=2$ outperformed $q=1$. However, we did not investigate the performance of higher values of $q$. We leave the quantification of the benefits associated with using higher values of $q$, as well as the extension of $q=\infty$ aligning with the supremum norm $L_\infty$, and an exploration of their advantages, for future research.

\newpage

\vspace*{1pc}

\noindent \textbf{Funding}\\
\noindent \textit{This work has been supported by the Research Training Group ``Biostatistical Methods for High-Dimensional Data in Toxicology'' (RTG 2624, Project P5) funded by the Deutsche Forschungsgemeinschaft (DFG, German Research Foundation - Project Number 427806116).}
\bigskip

\noindent \textbf{Conflict of Interest}

\noindent \textit{The authors have declared no conflict of interest.}
\bigskip

\noindent \textbf{Data Availability Statement}

\noindent The code that supports the findings of this study is openly available in Zenodo at~\citep{schuermeyer-2025-code-optim}.
\newpage

\section*{Appendix}
\subsection*{Tables}

\begin{table}[hb]
  \centering
  \caption{Considered designs with information of used dose combination levels and corresponding weights based on case study data (see Section \ref{Case study}).
    $K$ denotes the number of distinct dose combinations,
    and for the numerically obtained designs, $\mathrm{elb}$ is the lower bound on their efficiency from~\eqref{eq:elb} in Corollary~\ref{corollary-elb}.}
  \label{tab:Designs}
  \small
  \begin{tabular}{lll*{6}{c@{\hskip2pt}}}
    \toprule
    Design & $K$ & $1-\mathrm{elb}$ & \multicolumn{6}{c}{Doses and Corresponding Weights}\\
    \midrule
    \multirow{4}{4em}{Ray $4/2$} & \multirow{4}{*}{$8$} & \multirow{4}{*}{ } & $(0.0, 0.0)$ & $(0.0, 2.33)$
                                                        & $(0.0, 4.67)$ & $(0.0, 7.0)$ & $(6.67, 0.0)$ & $(13.33, 0.0)$ \\
           &   &   & $1/8$ & $1/8$ & $1/8$ & $1/8$ & $1/8$ & $1/8$ \\
           &   &   & $(20.0, 0.0)$ & $(20.0, 7.0)$ &   &   &   &   \\
           &   &   & $1/8$ & $1/8$ &   &   &   &   \\
    \midrule
    \multirow{4}{4em}{Ray $4/4$} & \multirow{4}{*}{$10$} & \multirow{4}{*}{ } & $(0.0, 0.0)$ & $(0.0, 2.33)$ & $(0.0, 4.67)$ & $(0.0, 7.0)$ & $(6.67, 0.0)$ & $(6.67, 2.33)$ \\
           &   &   & $1/10$ & $1/10$ & $1/10$ & $1/10$ & $1/10$ & $1/10$ \\
           &   &   & $(13.33, 0.0)$ & $(13.33, 4.67)$ & $(20.0, 0.0)$ & $(20.0, 7.0)$ &   &   \\
           &   &   & $1/10$ & $1/10$ & $1/10$ & $1/10$ &   &   \\
    \midrule
    \multirow{6}{4em}{Factorial $4\times 4$} & \multirow{6}{*}{$16$} & \multirow{6}{*}{ } & $(0.0, 0.0)$ & $(0.0, 2.33)$ & $(0.0, 4.67)$ & $(0.0, 7.0)$ & $(6.67, 0.0)$ & $(6.67, 2.33)$ \\
           &   &   & $1/16$ & $1/16$ & $1/16$ & $1/16$ & $1/16$ & $1/16$ \\
           &   &   & $(6.67, 4.67)$ & $(6.67, 7.0)$ & $(13.33, 0.0)$ & $(13.33, 2.33)$ & $(13.33, 4.67)$ & $(13.33, 7.0)$ \\
           &   &   & $1/16$ & $1/16$ & $1/16$ & $1/16$ & $1/16$ & $1/16$ \\
           &   &   & $(20.0, 0.0)$ & $(20.0, 2.33)$ & $(20.0, 4.67)$ & $(20.0, 7.0)$ &   &   \\
           &   &   & $1/16$ & $1/16$ & $1/16$ & $1/16$ &   &   \\
    \midrule
    \multirow{4}{4em}{Original} & \multirow{4}{*}{$12$} & \multirow{4}{*}{ } & $(0.0, 0.0)$ & $(5.0, 0.0)$ & $(10.0, 0.0)$ & $(20.0, 0.0)$ & $(0.0, 0.3)$ & $(0.0, 1.0)$ \\
           &   &   & $0.22$ & $0.11$ & $0.11$ & $0.16$ & $0.05$ & $0.05$ \\
           &   &   & $(0.0, 3.0)$ & $(10.0, 3.0)$ & $(20.0, 3.0)$ & $(0.0, 7.0)$ & $(5.0, 7.0)$ & $(20.0, 7.0)$ \\
           &   &   & $0.05$ & $0.05$ & $0.05$ & $0.09$ & $0.03$ & $0.03$ \\
    \midrule
    \multirow{4}{4em}{D-optimal} & \multirow{4}{*}{$9$} & \multirow{4}{*}{ } & $(0.0, 0.0)$ & $(0.0, 0.4)$ & $(0.0, 7.0)$ & $(0.0, 3.13)$ & $(3.66, 0.0)$ & $(3.8, 0.44)$ \\
           &   &   & $0.105$ & $0.099$ & $0.125$ & $0.125$ & $0.097$ & $0.073$ \\
           &   &   & $(7.67, 0.01)$ & $(20.0, 0.0)$ & $(20.0, 7.0)$ &   &   &   \\
           &   &   & $0.126$ & $0.125$ & $0.125$ &   &   &   \\
    \midrule
    \multirow{4}{4em}{MED $(10, 50)$} & \multirow{4}{*}{$11$} & \multirow{4}{*}{$\num{6.0e-5}$} & $(0.0, 0.0)$ & $(0.0, 0.3)$ & $(0.0, 7.0)$ & $(0.0, 2.61)$ & $(3.14, 0.0)$ & $(3.59, 1.99)$ \\
           &   &   & $0.049$ & $0.178$ & $0.012$ & $0.175$ & $0.107$ & $0.125$ \\
           &   &   & $(3.67, 0.48)$ & $(6.18, 0.38)$ & $(6.86, 0.0)$ & $(20.0, 0.0)$ & $(20.0, 7.0)$ &   \\
           &   &   & $0.107$ & $0.109$ & $0.115$ & $0.013$ & $0.01$ &   \\
    \midrule
    \multirow{4}{4em}{MED $(20, 80)$} & \multirow{4}{*}{$12$} & \multirow{4}{*}{$\num{3.3e-7}$} & $(0.0, 0.0)$ & $(0.0, 7.0)$ & $(0.0, 0.46)$ & $(1.16, 4.23)$ & $(3.27, 3.5)$ & $(3.55, 0.55)$ \\
           &   &   & $0.011$ & $0.051$ & $0.144$ & $0.168$ & $0.005$ & $0.099$ \\
           &   &   & $(3.66, 0.0)$ & $(7.57, 2.54)$ & $(9.15, 0.35)$ & $(9.4, 0.0)$ & $(20.0, 0.01)$ & $(20.0, 7.0)$ \\
           &   &   & $0.097$ & $0.151$ & $0.045$ & $0.057$ & $0.137$ & $0.035$ \\
    \bottomrule
  \end{tabular}
\end{table}

\begin{table}
  \centering
  \caption{Considered designs with information of used dose combination levels and corresponding weights for Scenario~2. $K$ denotes the number of distinct dose combinations,
    and for the numerically obtained designs, $\mathrm{elb}$ is the lower bound on their efficiency from~\eqref{eq:elb} in Corollary~\ref{corollary-elb}.}
  \label{tab:DesignsScenario2}
  \small
	\begin{tabular}{lll*{6}{c@{\hskip2pt}}}
	\toprule
	Design & $K$ & $1-\mathrm{elb}$ & \multicolumn{6}{c}{Doses and Corresponding Weights}\\
	\midrule
	\multirow{2}{5em}{Ray $3/2$} & \multirow{2}{*}{$6$} & \multirow{2}{*}{ } & $(0.0, 0.0)$ & $(0.0, 0.5)$ & $(0.0, 1.0)$ & $(0.5, 0.0)$ & $(1.0, 0.0)$ & $(1.0, 1.0)$ \\
	  &   &   & $1/6$ & $1/6$ & $1/6$ & $1/6$ & $1/6$ & $1/6$ \\
	\midrule
	\multirow{4}{5em}{Ray $4/2$} & \multirow{4}{*}{$8$} & \multirow{4}{*}{ } & $(0.0, 0.0)$ & $(0.0, 0.33)$ & $(0.0, 0.67)$ & $(0.0, 1.0)$ & $(0.33, 0.0)$ & $(0.67, 0.0)$ \\
	  &   &   & $1/8$ & $1/8$ & $1/8$ & $1/8$ & $1/8$ & $1/8$ \\
	  &   &   & $(1.0, 0.0)$ & $(1.0, 1.0)$ &   &   &   &   \\
	  &   &   & $1/8$ & $1/8$ &   &   &   &   \\
	\midrule
	\multirow{4}{5em}{Factorial $3\times 3$} & \multirow{4}{*}{$9$} & \multirow{4}{*}{ } & $(0.0, 0.0)$ & $(0.0, 0.5)$ & $(0.0, 1.0)$ & $(0.5, 0.0)$ & $(0.5, 0.5)$ & $(0.5, 1.0)$ \\       
	  &   &   & $1/9$ & $1/9$ & $1/9$ & $1/9$ & $1/9$ & $1/9$ \\
	  &   &   & $(1.0, 0.0)$ & $(1.0, 0.5)$ & $(1.0, 1.0)$ &   &   &   \\
	  &   &   & $1/9$ & $1/9$ & $1/9$ &   &   &   \\
	\midrule
	\multirow{6}{5em}{Factorial $4\times 4$} & \multirow{6}{*}{$16$} & \multirow{6}{*}{ } & $(0.0, 0.0)$ & $(0.0, 0.33)$ & $(0.0, 0.67)$ & $(0.0, 1.0)$ & $(0.33, 0.0)$ & $(0.33, 0.33)$ \\ 
	  &   &   & $1/16$ & $1/16$ & $1/16$ & $1/16$ & $1/16$ & $1/16$ \\
	  &   &   & $(0.33, 0.67)$ & $(0.33, 1.0)$ & $(0.67, 0.0)$ & $(0.67, 0.33)$ & $(0.67, 0.67)$ & $(0.67, 1.0)$ \\
	  &   &   & $1/16$ & $1/16$ & $1/16$ & $1/16$ & $1/16$ & $1/16$ \\
	  &   &   & $(1.0, 0.0)$ & $(1.0, 0.33)$ & $(1.0, 0.67)$ & $(1.0, 1.0)$ &   &   \\
	  &   &   & $1/16$ & $1/16$ & $1/16$ & $1/16$ &   &   \\
	\midrule
	\multirow{4}{5em}{MED $(10)$} & \multirow{4}{*}{$7$} & \multirow{4}{*}{$0$} & $(0.0, 0.18)$ & $(0.0, 0.58)$ & $(0.11, 0.09)$ & $(0.23, 0.0)$ & $(0.34, 0.28)$ & $(0.73, 0.0)$ \\
	  &   &   & $0.283$ & $0.07$ & $0.177$ & $0.323$ & $0.032$ & $0.071$ \\
	  &   &   & $(1.0, 1.0)$ &   &   &   &   &   \\
	  &   &   & $0.044$ &   &   &   &   &   \\
	\midrule
	\multirow{2}{5em}{D-optimal} & \multirow{2}{*}{$6$} & \multirow{2}{*}{ } & $(0.0, 0.25)$ & $(0.0, 0.6)$ & $(0.32, 0.0)$ & $(0.33, 0.26)$ & $(0.75, 0.0)$ & $(1.0, 1.0)$ \\
	  &   &   & $0.147$ & $0.152$ & $0.148$ & $0.2$ & $0.153$ & $0.2$ \\
	\bottomrule
	\end{tabular}
\end{table}

\begin{table}
  \centering
  \caption{Considered designs in robustness analyses setting 1 with information of used dose combination levels and corresponding weights.
    $K$ denotes the number of distinct dose combinations,
    and $\mathrm{elb}$ is the lower bound on their efficiency from~\eqref{eq:elb} in Corollary~\ref{corollary-elb}.
  }
  \label{tab:DesignsScenarioBayesian9095and1030}
  \small
  \begin{tabular}{lll*{6}{c@{\hskip2pt}}}
\toprule
Design & $K$ & $1-\mathrm{elb}$ & \multicolumn{6}{c}{Doses and Corresponding Weights}\\
\midrule
\multirow{4}{5em}{MED $(10, 30)$ $\gamma_t=0.02$} & \multirow{4}{*}{$9$} & \multirow{4}{*}{$\num{0.00021}$} & $(0.0, 12.0)$ & $(0.0, 0.0)$ & $(0.0, 4.87)$ & $(1.54, 3.18)$ & $(1.59, 12.0)$ & $(1.99, 0.0)$ \\
  &   &   & $0.187$ & $0.13$ & $0.061$ & $0.27$ & $0.027$ & $0.039$ \\
  &   &   & $(10.0, 3.02)$ & $(10.0, 12.0)$ & $(10.0, 0.0)$ &   &   &   \\
  &   &   & $0.032$ & $0.016$ & $0.238$ &   &   &   \\
\midrule
\multirow{4}{5em}{MED $(10, 30)$ $\gamma=0.005$} & \multirow{4}{*}{$7$} & \multirow{4}{*}{$\num{0.0034}$} & $(0.0, 0.0)$ & $(0.0, 12.0)$ & $(0.0, 3.97)$ & $(1.53, 3.04)$ & $(2.05, 0.0)$ & $(10.0, 12.0)$ \\
  &   &   & $0.139$ & $0.064$ & $0.23$ & $0.259$ & $0.22$ & $0.012$ \\
  &   &   & $(10.0, 0.0)$ &   &   &   &   &   \\
  &   &   & $0.076$ &   &   &   &   &   \\
\midrule
\multirow{4}{5em}{Bayesian MED $(10, 30)$} & \multirow{4}{*}{$9$} & \multirow{4}{*}{$\num{3.3e-8}$} & $(0.0, 0.0)$ & $(0.0, 12.0)$ & $(0.0, 3.82)$ & $(1.39, 12.0)$ & $(1.6, 3.26)$ & $(1.97, 0.0)$ \\
  &   &   & $0.121$ & $0.117$ & $0.18$ & $0.022$ & $0.207$ & $0.168$ \\
  &   &   & $(10.0, 0.0)$ & $(10.0, 2.8)$ & $(10.0, 12.0)$ &   &   &   \\
  &   &   & $0.146$ & $0.024$ & $0.015$ &   &   &   \\
\bottomrule
\end{tabular}

\end{table}

\begin{table}
  \centering
  \caption{Considered designs in robustness analyses setting 2 with information of used dose combination levels and corresponding weights.
  $K$ denotes the number of distinct dose combinations,
    and $\mathrm{elb}$ is the lower bound on their efficiency from~\eqref{eq:elb} in Corollary~\ref{corollary-elb}.}
  \label{tab:DesignsScenarioBayesianprior5}
  \small
  \begin{tabular}{lll*{6}{c@{\hskip2pt}}}
    \toprule
    Design & $K$ & $1-\mathrm{elb}$ & \multicolumn{6}{c}{Doses and Corresponding Weights}\\
    \midrule
    \multirow{4}{5em}{MED $(80, 90)$ $\gamma=0.02$} & \multirow{4}{*}{$7$} & \multirow{4}{*}{$\num{5.5e-7}$} & $(0.0, 0.0)$ & $(0.0, 12.0)$ & $(3.27, 12.0)$ & $(3.29, 5.92)$ & $(10.0, 0.0)$ & $(10.0, 5.88)$ \\
           &   &   & $0.004$ & $0.019$ & $0.313$ & $0.084$ & $0.02$ & $0.337$ \\
           &   &   & $(10.0, 12.0)$ &   &   &   &   &   \\
           &   &   & $0.223$ &   &   &   &   &   \\
    \midrule
    \multirow{4}{5em}{MED $(80, 90)$ $\gamma_t=-0.01$} & \multirow{4}{*}{$9$} & \multirow{4}{*}{$\num{4.0e-6}$} & $(0.0, 0.0)$ & $(0.0, 12.0)$ & $(0.0, 3.88)$ & $(1.74, 12.0)$ & $(1.94, 0.0)$ & $(2.34, 4.67)$ \\
           &   &   & $0.018$ & $0.107$ & $0.014$ & $0.167$ & $0.017$ & $0.233$ \\
           &   &   & $(10.0, 0.0)$ & $(10.0, 3.49)$ & $(10.0, 12.0)$ &   &   &   \\
           &   &   & $0.068$ & $0.218$ & $0.158$ &   &   &   \\
    \midrule
    \multirow{4}{5em}{Bayesian MED $(80, 90)$} & \multirow{4}{*}{$9$} & \multirow{4}{*}{$\num{0.0014}$} & $(0.0, 0.0)$ & $(0.0, 5.35)$ & $(0.0, 12.0)$ & $(2.16, 4.22)$ & $(2.32, 12.0)$ & $(2.68, 0.0)$ \\
           &   &   & $0.016$ & $0.05$ & $0.076$ & $0.109$ & $0.211$ & $0.036$ \\
           &   &   & $(10.0, 0.0)$ & $(10.0, 4.61)$ & $(10.0, 12.0)$ &   &   &   \\
           &   &   & $0.095$ & $0.227$ & $0.18$ &   &   &   \\
    \bottomrule
  \end{tabular}
\end{table}

\FloatBarrier

\clearpage
\subsection*{Figures}

\begin{figure}[b]
    \centering
    \includegraphics[width=0.8\linewidth]{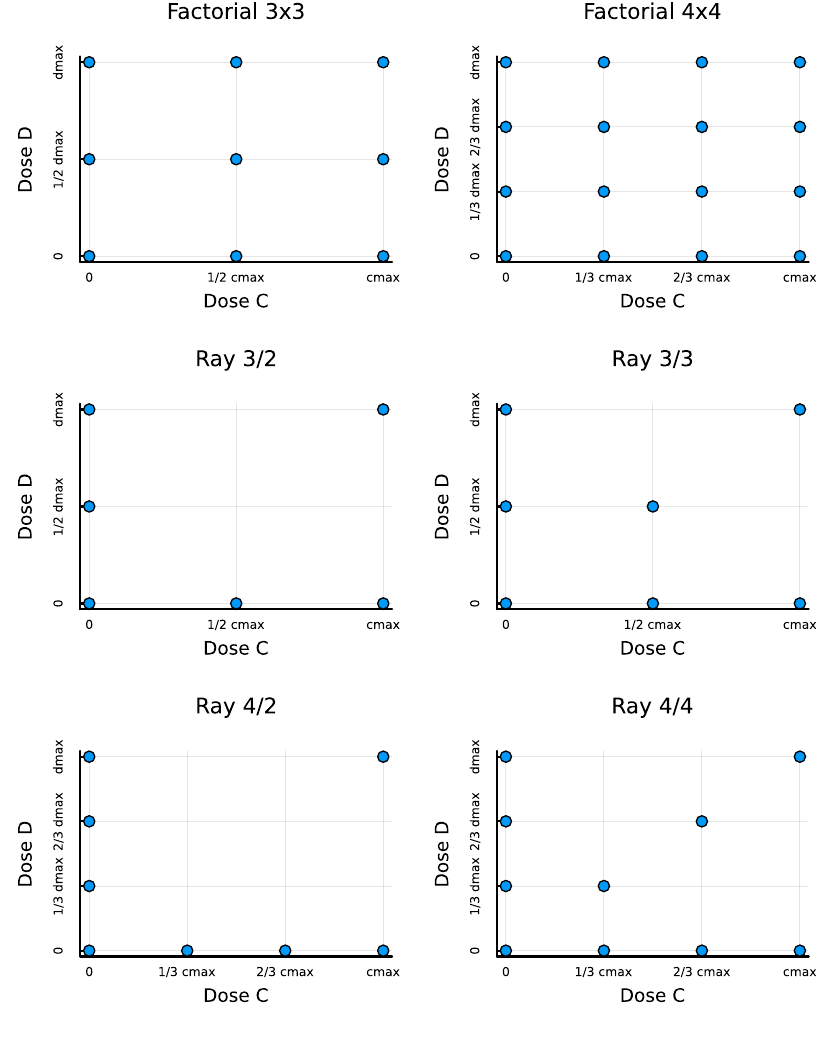}
    \caption{Showcase visualization of the considered factorial designs and ray designs. The considered number of support points can differ depending on the considered setting.}
    \label{fig:FactorialRayVisualization}
\end{figure}

\begin{figure}
    \centering
    \subfigure[Locally $\mathrm{MED}$-optimal design with $\mathrm{MED}_{10}$ set.]{\includegraphics[width=0.49\textwidth]{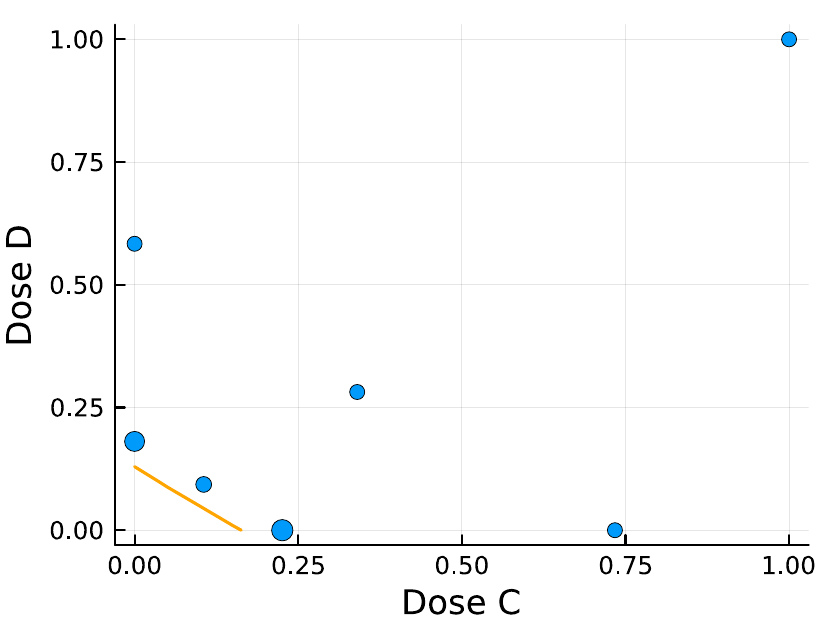}}
    \subfigure[D-optimal design with $\mathrm{MED}_{10}$ set.]{\includegraphics[width=0.49\textwidth]{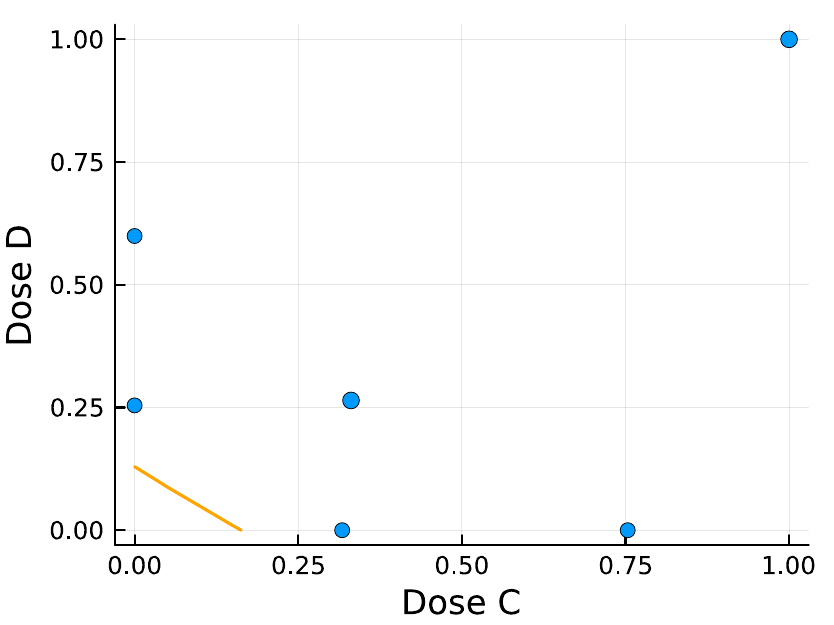}}
    \caption{Visualization of the considered designs and contour line in Scenario~2,
      which includes their support points represented by dots, with their sizes corresponding to their weights.}
    \label{fig: VisualizationDesignsScenario2}
\end{figure}

\begin{figure}
    \centering
    \includegraphics[width=1\linewidth]{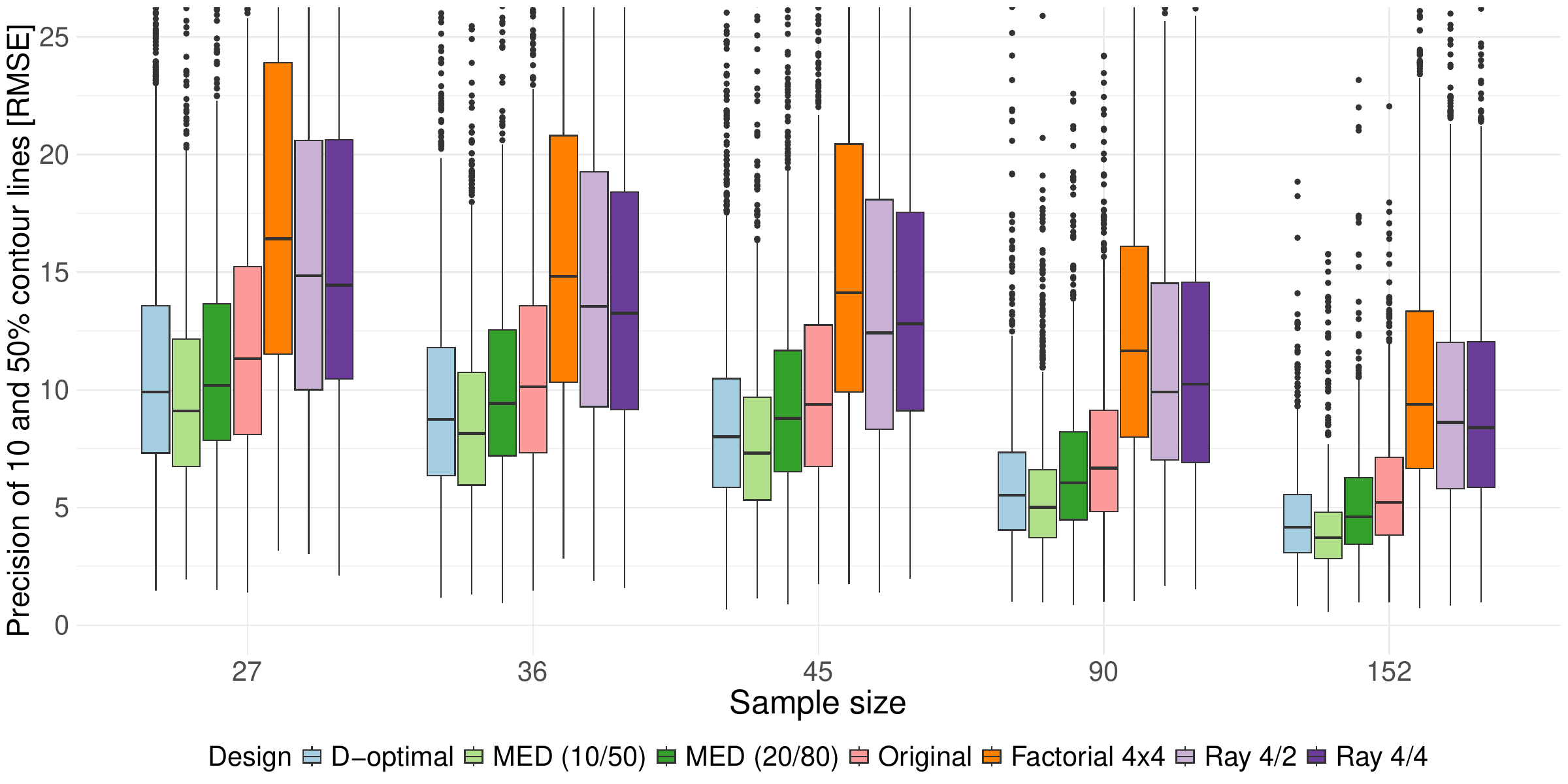}
    \caption{Extract of RMSE values grouped by design and total number of observations used in the simulation steps of Scenario~1. The locally $\mathrm{MED} (10,50)$-optimal design shows the smallest RMSE values for all considered sample sizes compared to all other designs, especially the traditionally used factorial design and both ray designs.
    }
    \label{fig:SImResults_BIData}
\end{figure}

\begin{figure}
    \centering
    \includegraphics[width=1\linewidth]{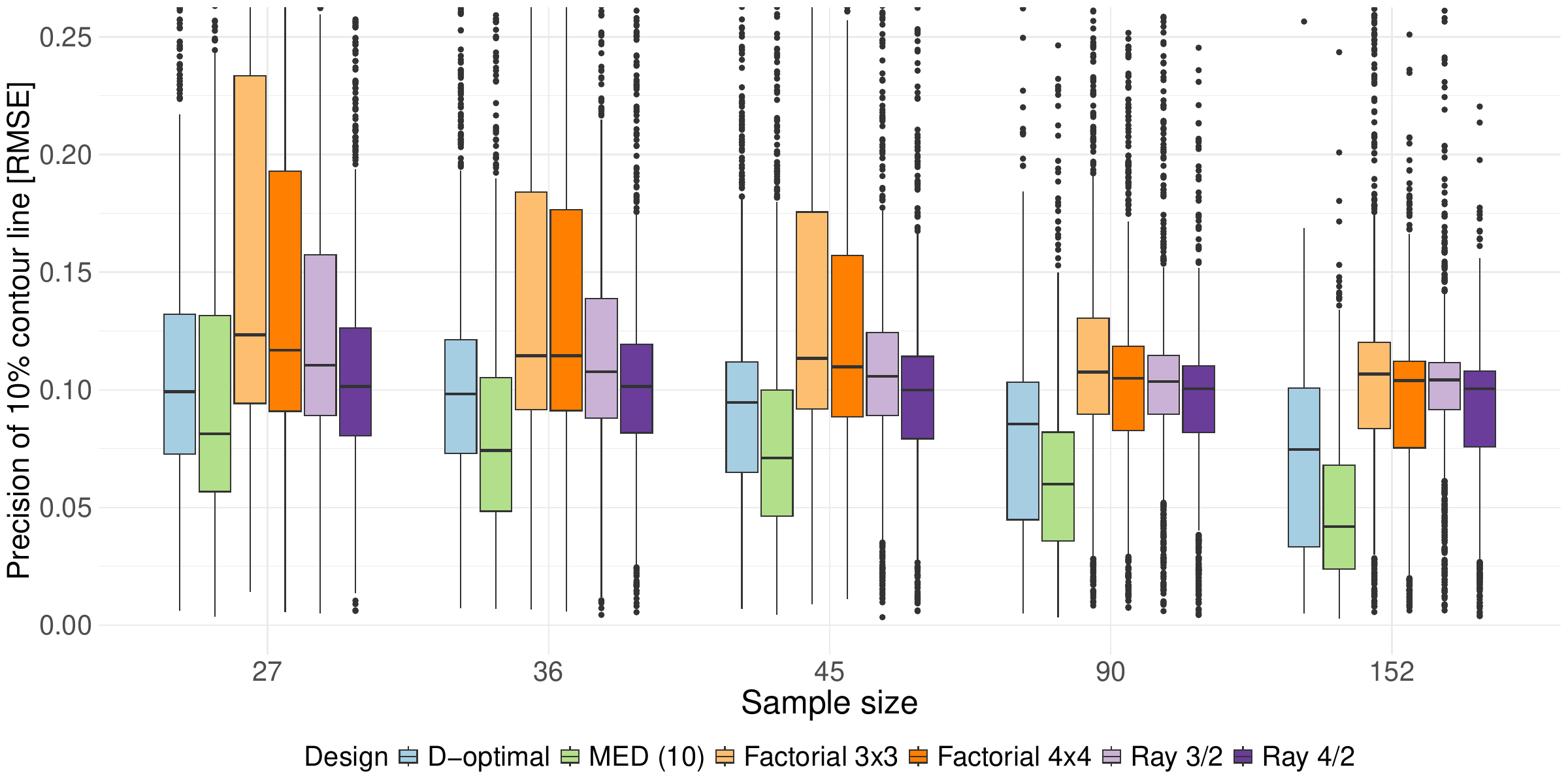}
    \caption{Extract of RMSE values grouped by design and total number of observations used in the simulation steps of Scenario~2. The locally $\mathrm{MED}$-optimal design outperforms all other designs for all numbers of measurements. In detail it shows the smallest RMSE-values and therefore the highest precision of the set of effective doses.
    }
    \label{fig:SImResults_Scenario2ALL}
\end{figure}

\begin{figure}
    \centering
    \includegraphics[width=1\linewidth]{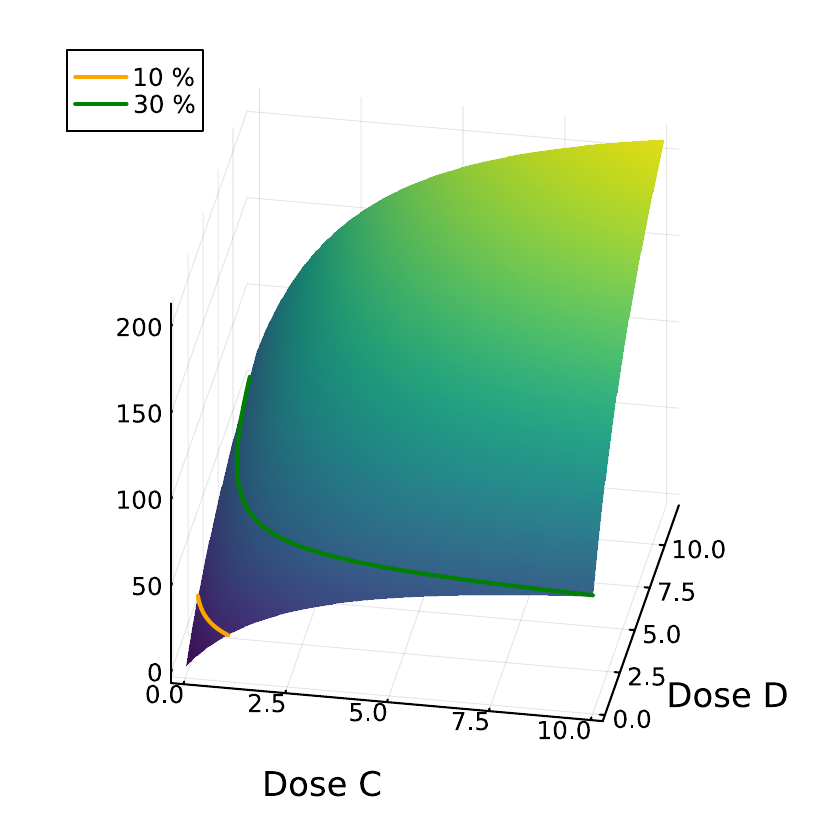}
    \caption{Additive response-surface model (including two Emax models) for robustness setting~1 with the considered 10\% and 30\% contour lines.
    }
    \label{fig:BayesianSetting1_OldSzenario2}
\end{figure}

\begin{figure}
    \centering
    \includegraphics[width=0.8\textwidth]{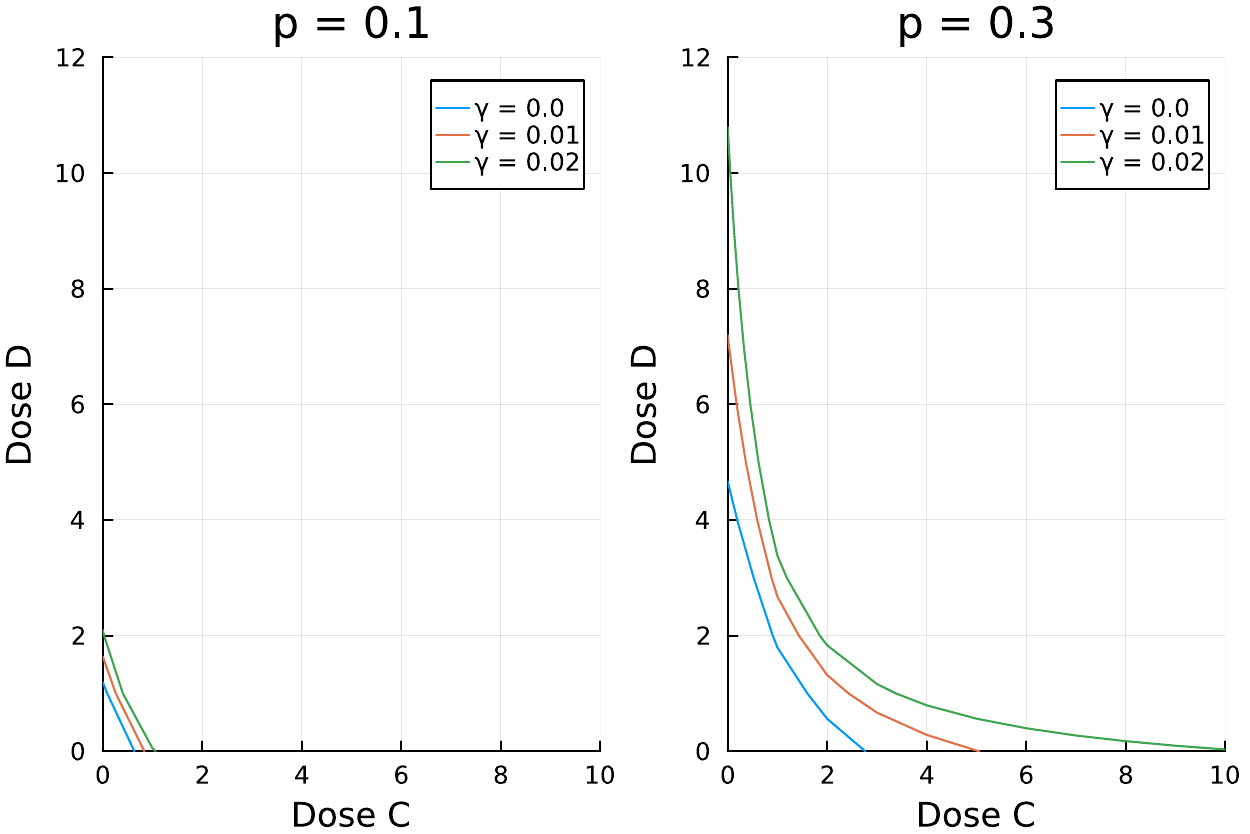}
    \caption{Visualization of $\mathrm{MED}_{10}$ and $\mathrm{MED}_{30}$ sets for robustness analysis setting 1 in case of different values of the interaction effect $\gamma$.
    }
    \label{fig:ContourlinesBayesianDesign9095}
\end{figure}

\begin{figure}
    \centering
    \subfigure[Locally $\mathrm{MED}$-optimal design with $\mathrm{MED}_{10}$ and $\mathrm{MED}_{30}$ sets for misspecified $\gamma=0.005$.]{\includegraphics[width=0.49\textwidth]{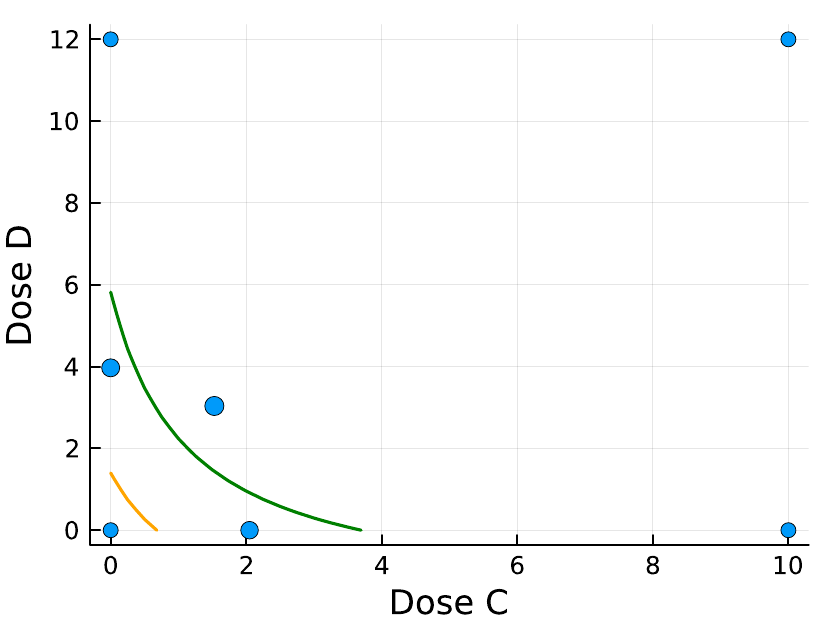}}
    \subfigure[Bayesian $\mathrm{MED}$-optimal design.]{\includegraphics[width=0.49\textwidth]{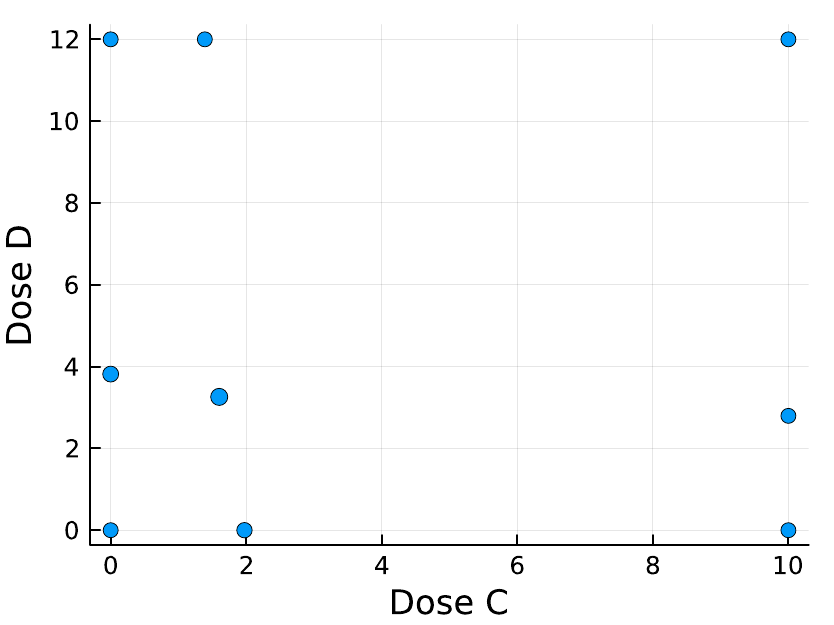}}
    \caption{Visualization of the considered designs in the robustness analysis setting 1, which includes their support points represented by dots, with their sizes corresponding to their weights.}
    \label{fig: VisualizationBayesianDesigns9095}
\end{figure}

\begin{figure}
    \centering
    \includegraphics[width=0.8\textwidth]{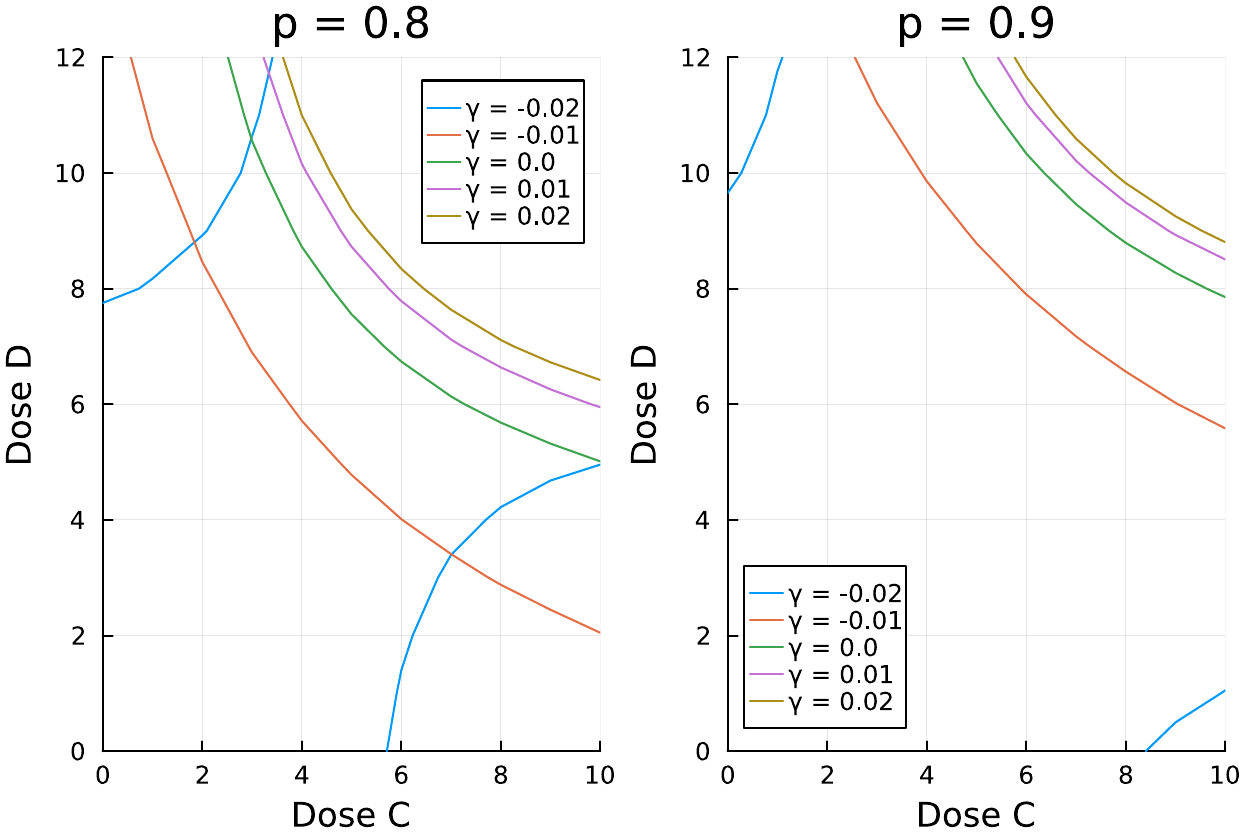}
    \caption{Visualization of $\mathrm{MED}_{80}$ and $\mathrm{MED}_{90}$ sets for the second setting in the robust analysis in case of different values of the interaction effect $\gamma$.
    }
    \label{fig:ContourlinesBayesianDesign_prior5}
\end{figure}

\begin{figure}[ht]
    \centering
    \subfigure[$\mathrm{MED}$ design with $\mathrm{MED}_{80}$ and $\mathrm{MED}_{90}$ sets for $\gamma_t=-0.01$.]{\includegraphics[width=0.49\textwidth]{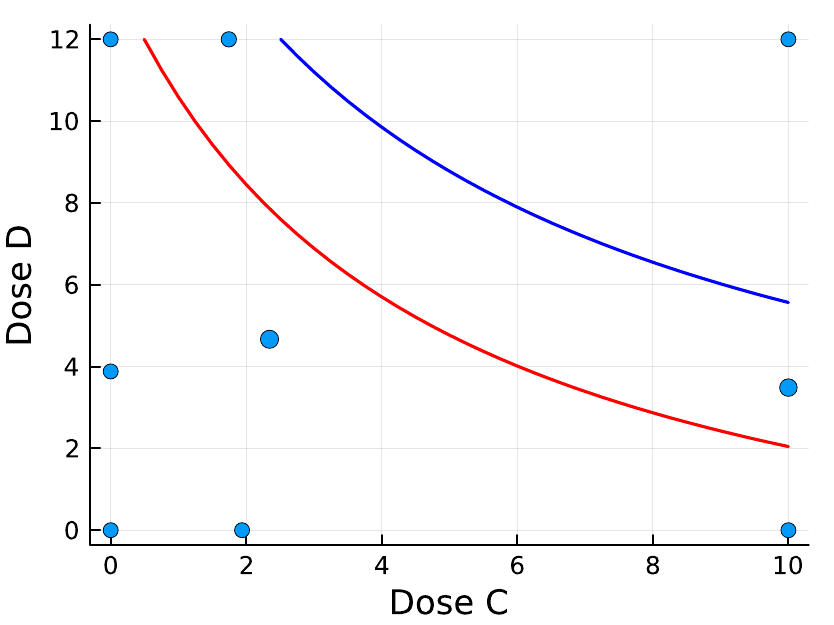}}
    \subfigure[Bayesian $\mathrm{MED}$ design.]{\includegraphics[width=0.49\textwidth]{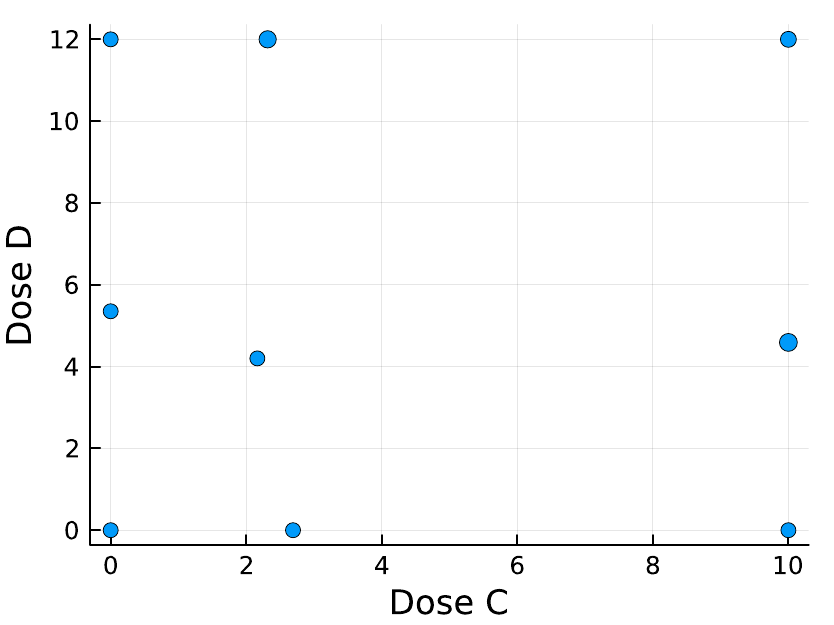}}
    \caption{Visualization of the considered Bayesian designs in robustness analysis setting 2,
      which includes their support points represented by dots, with their sizes corresponding to their weights.}
    \label{fig: VisualizationBayesianDesigns_prior5}
\end{figure}

\FloatBarrier

\bibliographystyle{abbrvnat}
\bibliography{references}

@article{ChalonerLarntz,
  author =       "Chaloner, Kathryn and Larntz, Kinley",
  title =        {Optimal {Bayesian} Design Applied To Logistic
                  Regression Experiments},
  journal =      "Journal of Statistical Planning and Inference",
  volume =       21,
  number =       2,
  pages =        {191-208},
  year =         1989,
  doi =          {10.1016/0378-3758(89)90004-9},
}

@manual{RCoreTeam2024,
  address =      {Vienna, Austria},
  author =       {{R Core Team}},
  organization = {{R} Foundation for Statistical Computing},
  title =        {{R}: A Language and Environment for Statistical
                  Computing},
  url =          {https://www.R-project.org/},
  year =         2024,
}

@article{SandigKirstine,
  author =       {Ludger Sandig},
  title =        {{Kirstine.jl}: A {Julia} Package for {Bayesian} Optimal
                  Design of Experiments},
  journal =      {Journal of Open Source Software},
  volume =       9,
  number =       97,
  pages =        6424,
  year =         2024,
  doi =          {10.21105/joss.06424},
}

@article{almohaimeed2014experimental,
  author =       {Almohaimeed, Bader and Donev, Alexander N},
  title =        {Experimental Designs for Drug Combination Studies},
  journal =      {Computational Statistics \& Data Analysis},
  volume =       71,
  pages =        {1077-1087},
  year =         2014,
  doi =          {10.1016/j.csda.2013.01.007},
}

@article{bliss1939toxicity,
  author =       {Bliss, Chester I},
  title =        {The Toxicity of Poisons Applied Jointly},
  journal =      {Annals of Applied Biology},
  volume =       26,
  number =       3,
  pages =        {585-615},
  year =         1939,
  doi =          {10.1111/j.1744-7348.1939.tb06990.x},
}

@article{bornkamp2009mcpmod,
  author =       {Bornkamp, Bj{\"o}rn and Pinheiro, Jos{\'e} and
                  Bretz, Frank},
  title =        {{MCPMod}: an {R} Package for the Design and Analysis
                  of Dose-Finding Studies},
  journal =      {Journal of Statistical Software},
  volume =       29,
  pages =        {1-23},
  year =         2009,
  doi =          {10.18637/jss.v029.i07},
}

@article{bretz2005combining,
  author =       {Bretz, Frank and Pinheiro, Jos{\'e} C and Branson,
                  Michael},
  title =        {Combining Multiple Comparisons and Modeling
                  Techniques in Dose-Response Studies},
  journal =      {Biometrics},
  volume =       61,
  number =       3,
  pages =        {738-748},
  year =         2005,
  doi =          {10.1111/j.1541-0420.2005.00344.x},
}

@article{bretz2010practical,
  author =       {Bretz, Frank and Dette, Holger and Pinheiro, Jose C},
  title =        {Practical Considerations for Optimal Designs in
                  Clinical Dose Finding Studies},
  journal =      {Statistics in Medicine},
  volume =       29,
  number =       {7-8},
  pages =        {731-742},
  year =         2010,
  doi =          {10.1002/sim.3802},
}

@article{chaloner1989bayesian,
  author =       {Chaloner, Kathryn},
  title =        {{Bayesian} Design for Estimating the Turning Point
                  of a Quadratic Regression},
  journal =      {Communications in Statistics-Theory and Methods},
  volume =       18,
  number =       4,
  pages =        {1385-1400},
  year =         1989,
}

@article{chou2008preclinical,
  author =       {Chou, Ting-Chao},
  title =        {Preclinical Versus Clinical Drug Combination
                  Studies},
  journal =      {Leukemia \& Lymphoma},
  volume =       49,
  number =       11,
  pages =        {2059-2080},
  year =         2008,
  doi =          {10.1080/10428190802353591},
}

@article{dette2016optimal,
  author =       {Dette, Holger and Schorning, Kirsten},
  title =        {Optimal Designs for Comparing Curves},
  journal =      {Annals of Statistics},
  volume =       44,
  number =       3,
  pages =        1103,
  year =         2016,
  doi =          {10.1214/15-AOS1399},
}

@book{fedorov2013optimal,
  author =       {Fedorov, Valerii V and Leonov, Sergei L},
  title =        {Optimal design for nonlinear response models},
  year =         2013,
  publisher =    {CRC Press},
  doi =          {10.1201/b15054},
}

@article{foucquier2015analysis,
  author =       {Foucquier, Julie and Guedj, Mickael},
  title =        {Analysis of Drug Combinations: Current
                  Methodological Landscape},
  journal =      {Pharmacology Research \& Perspectives},
  volume =       3,
  number =       3,
  year =         2015,
  doi =          {10.1002/prp2.149},
}

@article{hollandletz2018,
  author =       {Holland-Letz, Tim and Kopp-Schneider, Annette},
  title =        {Optimal Experimental Designs for Estimating the Drug
                  Combination Index in Toxicology},
  journal =      {Computational Statistics \& Data Analysis},
  volume =       117,
  pages =        {182-193},
  year =         2018,
  doi =          {10.1016/j.csda.2017.08.006},
}

@article{jennrich1969asymptotic,
  author =       {Jennrich, Robert I},
  title =        {Asymptotic Properties of Non-Linear Least Squares
                  Estimators},
  journal =      {The Annals of Mathematical Statistics},
  volume =       40,
  number =       2,
  pages =        {633-643},
  year =         1969,
  doi =          {10.1214/aoms/1177697731},
}

@article{jiang2014summarizing,
  author =       {Jiang, Xiaoqi and Kopp-Schneider, Annette},
  title =        {Summarizing {EC50} Estimates From Multiple
                  Dose-Response Experiments: a Comparison of a
                  Meta-Analysis Strategy To a Mixed-Effects Model
                  Approach},
  journal =      {Biometrical Journal},
  volume =       56,
  number =       3,
  pages =        {493-512},
  year =         2014,
  doi =          {10.1002/bimj.201300123},
}

@article{kiefer,
  author =       {Kiefer, J.},
  title =        {General Equivalence Theory for Optimum Designs
                  (Approximate Theory)},
  journal =      {The Annals of Statistics},
  volume =       2,
  number =       5,
  pages =        {849-879},
  year =         1974,
  doi =          {10.1214/aos/1176342810},
}

@article{lee2007interaction,
  author =       {Lee, J Jack and Kong, Maiying and Ayers, Gregory D
                  and Lotan, Reuben},
  title =        {Interaction Index and Different Methods for
                  Determining Drug Interaction in Combination Therapy},
  journal =      {Journal of Biopharmaceutical Statistics},
  volume =       17,
  number =       3,
  pages =        {461-480},
  year =         2007,
  doi =          {10.1080/10543400701199593},
}

@article{loewe1927mischarznei,
  author =       {Loewe, S},
  title =        {{Die Mischarznei: Versuch einer Allgemeinen
                  Pharmakologie der Arzneikombinationen}},
  journal =      {Klinische Wochenschrift},
  volume =       6,
  number =       23,
  pages =        {1077-1085},
  year =         1927,
  doi =          {https://doi.org/10.1007/BF01890305},
}

@InProceedings{lorensen-1987-march,
  author =       {Lorensen, William E. and Cline, Harvey E.},
  title =        {Marching cubes: A high resolution {3D} surface
                  construction algorithm},
  booktitle =    {Proceedings of the 14th annual conference on
                  Computer graphics and interactive techniques},
  year =         1987,
  pages =        {163-169},
  doi =          {10.1145/37401.37422},
  publisher =    {ACM},
  series =       {SIGGRAPH '87},
}

@article{masoudi2019metaheuristic,
  author =       {Masoudi, Ehsan and Holling, Heinz and Duarte,
                  Belmiro PM and Wong, Weng Kee},
  title =        {A Metaheuristic Adaptive Cubature Based Algorithm To
                  Find {Bayesian} Optimal Designs for Nonlinear
                  Models},
  journal =      {Journal of Computational and Graphical Statistics},
  volume =       28,
  number =       4,
  pages =        {861-876},
  year =         2019,
  doi =          {10.1080/10618600.2019.1601097},
}

@article{miller2007optimal,
  author =       {Miller, Frank and Guilbaud, Olivier and Dette,
                  Holger},
  title =        {Optimal Designs for Estimating the Interesting Part
                  of a Dose-Effect Curve},
  journal =      {Journal of Biopharmaceutical Statistics},
  volume =       17,
  number =       6,
  pages =        {1097-1115},
  year =         2007,
  doi =          {10.1080/10543400701645140},
}

@article{mokhtari2017combination,
  author =       {Mokhtari, Reza Bayat and Homayouni, Tina S and
                  Baluch, Narges and Morgatskaya, Evgeniya and Kumar,
                  Sushil and Das, Bikul and Yeger, Herman},
  title =        {Combination Therapy in Combating Cancer},
  journal =      {Oncotarget},
  volume =       8,
  number =       23,
  pages =        {38022-38043},
  year =         2017,
  doi =          {10.18632/oncotarget.16723},
}

@article{papathanasiou2019optimizing,
  author =       {Papathanasiou, Theodoros and Strathe, Anders and
                  Overgaard, Rune Viig and Lund, Trine Meldgaard and
                  Hooker, Andrew C},
  title =        {Optimizing Dose-Finding Studies for Drug
                  Combinations Based on Exposure-Response Models},
  journal =      {The AAPS Journal},
  volume =       21,
  pages =        {1-11},
  year =         2019,
  doi =          {10.1208/s12248-019-0365-3},
}

@article{pronzato1985robust,
  author =       {Pronzato, Luc and Walter, Eric},
  title =        {Robust Experiment Design Via Stochastic
                  Approximation},
  journal =      {Mathematical Biosciences},
  volume =       75,
  number =       1,
  pages =        {103-120},
  year =         1985,
  doi =          {10.1016/0025-5564(85)90068-9},
}

@book{pukelsheim,
  author =       "Pukelsheim, Friedrich",
  title =        "Optimal Design of Experiments",
  year =         2006,
  publisher =    "SIAM",
  address =      "Philadelphia",
  doi =          {10.1137/1.9780898719109},
}

@article{pukelsheimrieder1992,
  author =       {Friedrich Pukelsheim and Sabine Rieder},
  title =        {Efficient Rounding of Approximate Designs},
  journal =      {Biometrika},
  volume =       79,
  number =       4,
  pages =        {763-770},
  year =         1992,
  doi =          {10.2307/2337232},
}

@article{ronneberg2021bayesynergy,
  author =       {R{\o}nneberg, Leiv and Cremaschi, Andrea and Hanes,
                  Robert and Enserink, Jorrit M and Zucknick, Manuela},
  title =        {{bayesynergy}: Flexible {Bayesian} Modelling of
                  Synergistic Interaction Effects in in Vitro Drug
                  Combination Experiments},
  journal =      {Briefings in Bioinformatics},
  volume =       22,
  number =       6,
  year =         2021,
  doi =          {10.1093/bib/bbab251},
}

@misc{schuermeyer-2025-code-optim,
  author =       {Schürmeyer, Leonie and Sandig, Ludger and Kühne, Jorrit},
  doi =          {10.5281/zenodo.17865625},
  title =        {Code for ``{Optimal} designs for identifying effective
                  doses in drug combination studies'' (version 2)},
  year =         2025,
  note =         {doi: 10.5281/zenodo.17865625}
}

@article{straetemans2005design,
  author =       {Straetemans, Roel and O'Brien, Timothy and Wouters,
                  Luc and Van Dun, Jacky and Janicot, Michel and
                  Bijnens, Luc and Burzykowski, Tomasz and Aerts,
                  Marc},
  title =        {Design and Analysis of Drug Combination Experiments},
  journal =      {Biometrical Journal},
  volume =       47,
  number =       3,
  pages =        {299-308},
  year =         2005,
  doi =          {10.1002/bimj.200410124},
}

@article{zhao2010comparison,
  author =       {Zhao, Liang and Au, Jessie LS and Wientjes, M
                  Guillaume},
  title =        {Comparison of Methods for Evaluating Drug-Drug
                  Interaction},
  journal =      {Frontiers in Bioscience (Elite Edition)},
  volume =       2,
  pages =        241,
  year =         2010,
  doi =          {10.2741/e86},
}

@article{zhou2024combination,
title = {Combination {MCP}-Mod for two-drug combination dose-ranging studies},
doi = {10.1080/10543406.2024.2311254},
volume = 35,
    number = 2,
    pages = {257--270},
journal = {Journal of Biopharmaceutical Statistics},
author = {Zhou, Yifan and Sloan, Abigail and Menon, Sandeep and Wang, Ling},
year = 2024,
}

@article{liou2015response,
  title={Response surface models in the field of anesthesia: a crash course},
doi = {10.1016/j.aat.2015.06.005},
  author={Liou, Jing-Yang and Tsou, Mei-Yung and Ting, Chien-Kun},
  journal={Acta Anaesthesiologica Taiwanica},
  volume={53},
  number={4},
  pages={139--145},
  year={2015},
  publisher={Elsevier}
}

@incollection{macdougall2006analysis,
	title = {{Analysis of dose–response studies—Emax model}},
	pages = {127--145},
	booktitle = {Dose finding in drug development},
	publisher = {Springer},
	author = {Macdougall, James},
	year = {2006},
}

@article{chen-2022-partic-swarm,
  author =       {Chen, Ping-Yang and Chen, Ray-Bing and Wong, Weng
                  Kee},
  title =        {Particle Swarm Optimization for Searching Efficient
                  Experimental Designs: a Review},
  journal =      {WIREs Computational Statistics},
  volume =       14,
  number =       5,
  pages =        {e1578},
  year =         2022,
  doi =          {10.1002/wics.1578},
}

@article{yang-2013-optim-desig,
  author =       {Yang, Min and Biedermann, Stefanie and Tang, Elina},
  title =        {On Optimal Designs for Nonlinear Models: a General
                  and Efficient Algorithm},
  journal =      {Journal of the American Statistical Association},
  volume =       108,
  number =       504,
  pages =        {1411-1420},
  year =         2013,
  doi =          {10.1080/01621459.2013.806268},
  month =        {Dec},
  publisher =    {Informa UK Limited},
}

\includepdf[pages=-]{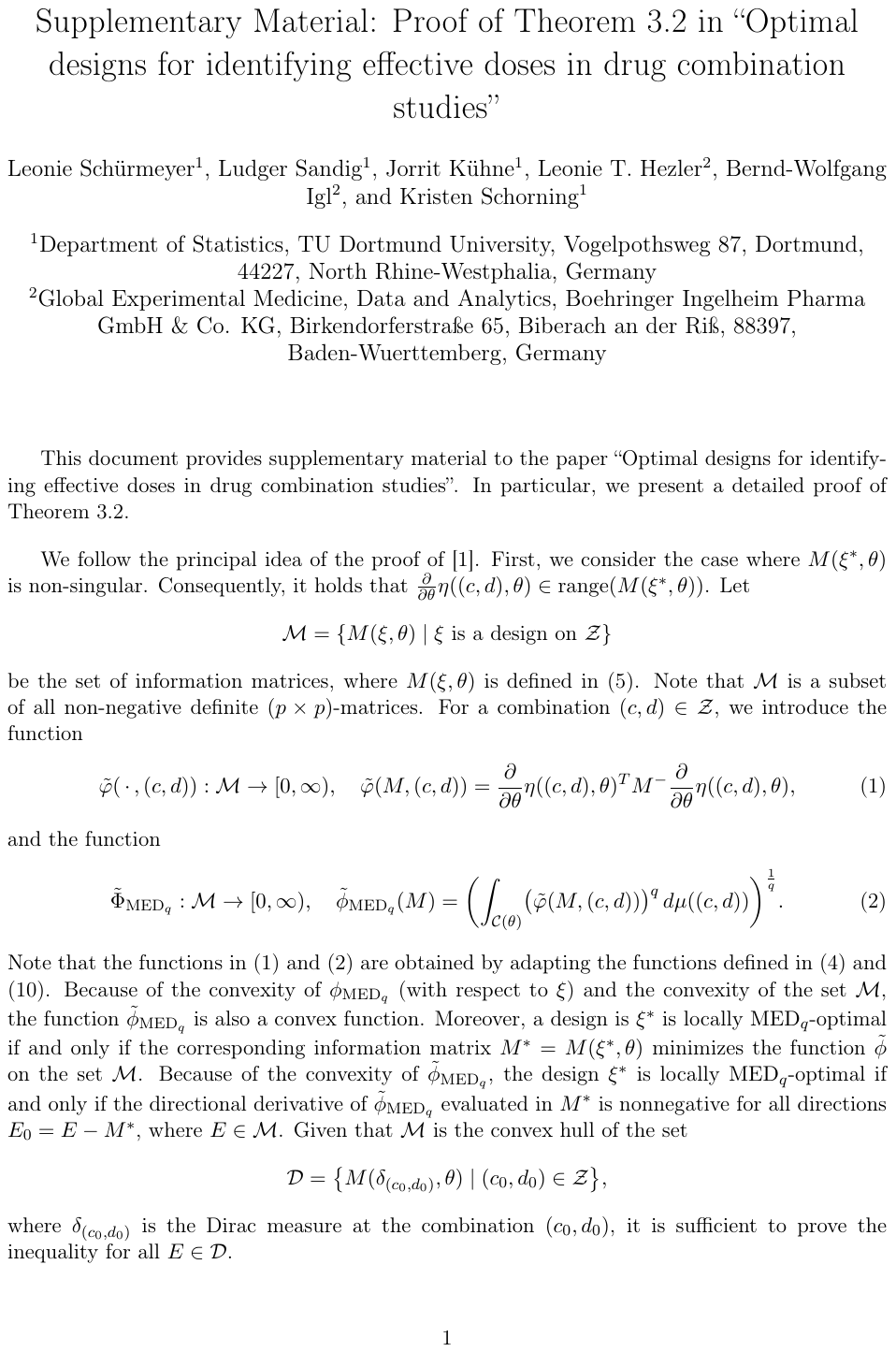}

\end{document}